\documentclass[3p, twocolumn]{elsarticle}

\usepackage{natbib}
\setcitestyle{authoryear}
\setcitestyle{round}
\usepackage{color}
\usepackage{geometry}
\usepackage{graphicx}
\usepackage{epsfig}
\usepackage{subcaption}
\usepackage{multirow} 
\usepackage{dcolumn}
\usepackage[fleqn]{amsmath}
\usepackage{amssymb}
\usepackage{bm}

\usepackage{upgreek}
\usepackage{textgreek}
\usepackage{hyperref}

\hypersetup{
  colorlinks,
  citecolor=Cyan,
  linkcolor=Blue,
  urlcolor=Pink}

\title{Fractional Quantum Hall Effect at the Filling Factor $\nu=5/2$}

\author{Ken K. W. Ma}
\address{National High Magnetic Field Laboratory, Tallahassee, Florida 32310, USA}
\ead{ken_ma@magnet.fsu.edu}

\author{Michael R. Peterson}
\address{Department of Physics $\&$ Astronomy, California State University Long Beach, Long Beach, California 90840, USA}

\author{V. W. Scarola}
\address{Department of Physics, Virginia Tech, Blacksburg, Virginia 24061, USA}

\author{Kun Yang}
\address{Department of Physics and National High Magnetic Field Laboratory, Florida State University, Tallahassee, Florida 32306, USA}
\ead{kunyang@magnet.fsu.edu}

\begin{document}

\begin{abstract}
The fractional quantum Hall  (FQH) effect at the filling factor $\nu=5/2$ was discovered in GaAs heterostructures more than 35 years ago. Various topological orders have been proposed as possible candidates to describe this FQH state. Some of them possess non-Abelian anyon excitations, an entirely new type of quasiparticle with fascinating properties.  If observed, non-Abelian anyons could offer fundamental building blocks of a topological quantum computer. Nevertheless, the nature of the FQH state at $\nu=5/2$ is still under debate. In this chapter, we provide an overview of the theoretical background, numerical results, and experimental measurements pertaining to this special FQH state. Furthermore, we review some recent developments and their possible interpretations. Possible future directions toward resolving the nature of the $5/2$ state are also discussed.
\end{abstract}


\maketitle

\section*{Notations and acronyms}

\noindent $\nu=\rho\phi_0/B$: Landau level filling factor\\
 $\ell_0=\sqrt{\hbar/eB}$: magnetic length\\
 $w$: quantum well thickness\\
 $\phi_0=h/e$: magnetic flux quantum\\
 $\nu$: electron filling factor\\
 $E_\mathrm{Coul} = e^2/\epsilon\ell_0$: Coulomb energy\\
 $\hbar\omega_c=eB/mc$: cyclotron energy\\
 $\mathcal{P}_\mathrm{SLL}$: projection operator to the second electronic Landau level\\
 $J=\prod_{1\leq i<j\leq N}(z_i-z_j)$: Jastrow factor\\
 $z_k=x_k-iy_k$: position of $k$-th electron in the $xy$-plane\\
 $\mathcal{C}$: Chern number\\
 $N$: number of electrons \\
 $Q$: monopole strength in the spherical geometry\\
 $S$: topological shift\\
 $\kappa = E_\mathrm{Coul}/\hbar\omega_c$: Landau level mixing parameter\\
 $K_H$: thermal Hall conductance \\
 CFT: conformal field theory\\
 FQH: fractional qauntum Hall \\
 QPC: quantum point contact\\
 SET: single electron transistor\\
 LL: Landau level\\
 DMRG: density matrix renormalization group
 
\section*{Key points and objective}

\begin{itemize}
\item The 5/2 fractional quantum Hall state was discovered more than 35 years ago but the nature of the physics underlying the state remains under intense debate.
     
\item The exciting possibility of a paired 5/2 state hosting exotic excitations with non-Abelian braid statistics led to considerable effort to prove their existence.  
     
\item Strong numerical and experimental evidence suggests that the 5/2 state arises from pairing between composite fermions but the precise nature of the topological order defining the paired state remains unsettled.
     
\item We review the culmination of experimental and theoretical studies on the 5/2 state and discuss possible future directions for further inquiry. 
 \end{itemize}

\section{Introduction}

The discovery of the FQH effect in two-dimensional semiconductor heterostructures started a new chapter in condensed matter physics \cite{Tsui1982}. For a recent review on the basic concepts in FQH physics, see \cite{Papic-review}. In contrast to the conventional wisdom from Landau-Ginzburg theory, different quantum Hall states cannot be distinguished by spontaneously broken symmetries~\cite{Wen_book}. Furthermore, early theoretical work suggested that the low-energy excitations in a FQH liquid are neither bosons nor fermions. Instead, they possess fractional charges~\cite{Laughlin} and fractional braiding statistics~\cite{Arovas1984, Halperin_hierarchy} [see also the review article by~\cite{Dima-fractional}], which are known as anyons~\cite{Leinaas1977, Wilczek-anyon}. This special type of excitation turns out to be the basic ingredient in the formulation of topological order, which provides an effective description of different FQH states and distinguishes them by specifying the sets of possible anyons the FQH states can support ~\cite{Wen_book}. Note that the existence of fractional statistics has received strong support from very recent experiments~\cite{Bartolomei, Nakamura2020, Heiblum-MZI}.

\begin{figure} [htb]
\centering
\includegraphics[width=3.0in]{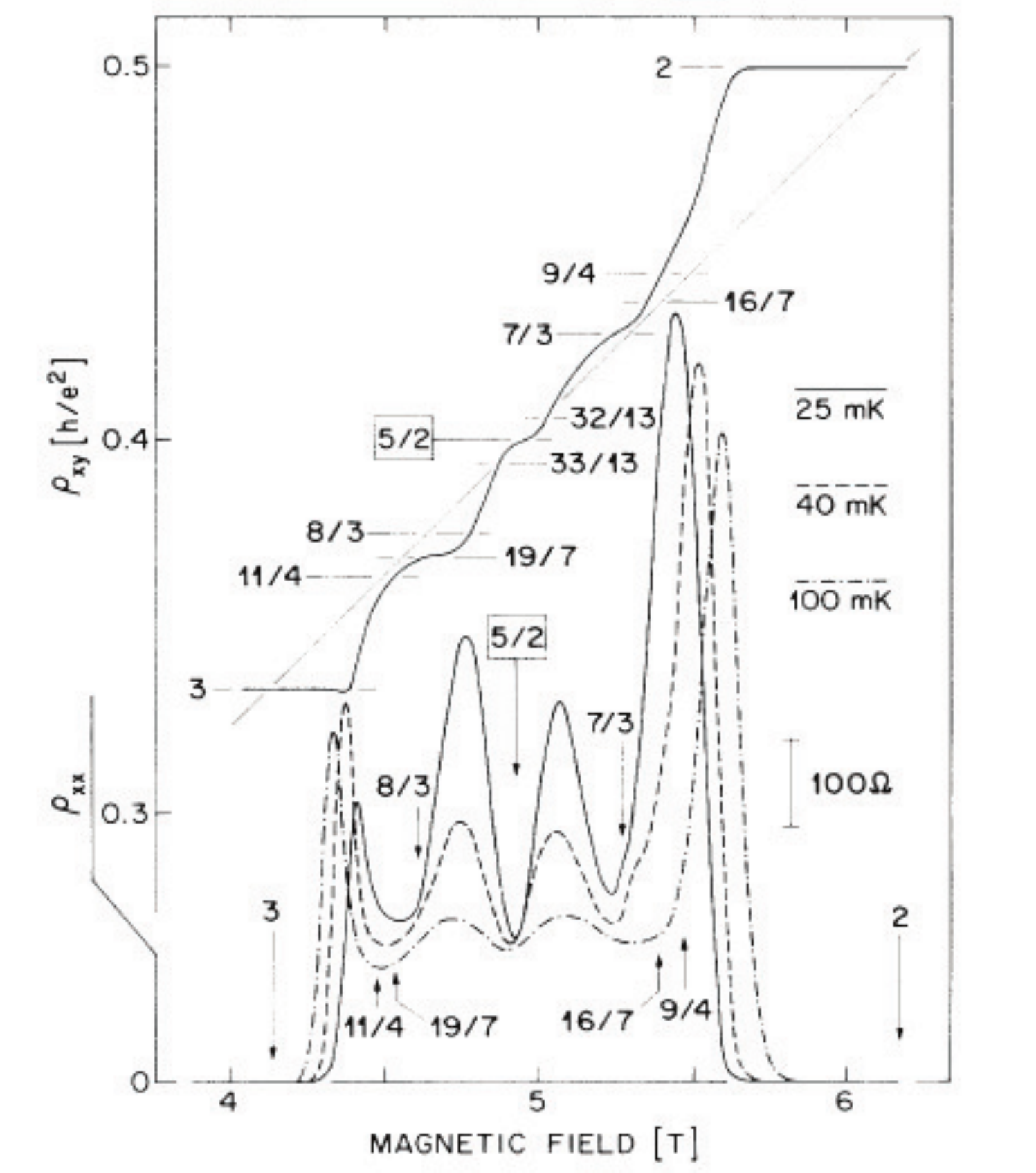}
\caption{The first experimental observation of the FQH state at $\nu=5/2$ revealed by plotting both the Hall ($\rho_{xy}$) and longitudinal ($\rho_{xx}$) resistivity versus perpendicular magnetic field. The figure is taken from~\cite{Willett1987} with permission.}
\label{fig:Willett-52}
\end{figure}

Many FQH states at different filling factors have been observed in various materials. Nearly all of them have odd denominators. For a summary of the observed FQH states, readers may refer to the lists in~\cite{FQH19/8_2} and \cite{Hidalgo2022}. Among them, the FQH state at filling factor $\nu=5/2$ (dubbed as the $5/2$ state below) in GaAs-GaAlAs quantum wells has attracted a tremendous amount of attention since its discovery~\cite{Willett1987}. This historical result is shown in Fig.~\ref{fig:Willett-52}. The observation of the $5/2$ state was surprising. It is a FQH state in the second LL (we define the lowest LL as the first LL). Also, it was the first FQH state with an even-denominator filling factor being observed in a single layer system.

The $5/2$ state has been challenging to understand, as compared to other FQH states in the lowest LL which usually have odd-denominator filling factors.  In particular, entirely new topological orders have been proposed as possible candidates to describe this special FQH state. Interestingly, some of them host non-Abelian anyons which may be useful in topological quantum computation~\cite{DasSarma2005, Kitaev2003, Das-Sarma}. 

Different types of experiments have been proposed and performed to probe the nature of the $5/2$ state. Meanwhile, numerical simulation is another essential tool to study the problem. Evidence from both sources strongly suggests that the $5/2$ state possesses non-Abelian topological order. Nevertheless, which one (or perhaps what combination of topological orders) is actually realized in real samples is still under considerable debate. A new twist on this 35-year old problem arose thanks to the exciting yet puzzling results from recent experiments~\cite{Banerjee2018, Heiblum-interface2, Heiblum-interface} that we will discuss below. In order to resolve the apparent conflict between the recent experimental and existing numerical results, various scenarios have been proposed~\cite{Mulligan2020, Dima_comment2018, Lian2018, Ken2019, Mross2018, Simon_PRB2018, Simon-reply, Simon2020, Wang2018}. 

In this chapter, we provide an overview of the theoretical background, numerical studies, and experimental findings pertaining to the $5/2$ state in GaAs heterostructures. Previous review articles on the same topic can be found in~\cite{5/2-review-2019, Lin-review, Peterson-review, Schreiber2020, Willett-review}. We will highlight recent developments in the past five years and outline possible future directions. Before diving into the details of the $5/2$ state, it is worthwhile to mention that FQH states at different half-integer filling factors have been observed in GaAs heterostructures and other materials. They are summarized in Table~\ref{tab:half}. For a recent review on half-integer FQH states in graphene-based systems, see~\cite{Charkaborty-review}. Note that an anomalous quantized plateau at $\nu=3/2$ has been observed in monolayer GaAs systems with gate-defined confined regions~\cite{Fu2019, Hayafuchi2022, Lin-3/2}. However, this plateau is believed to emerge from a larger filling factor in the bulk. Thus, its nature might be very different from a genuine half-integer FQH state.

We note the present chapter is different from (most) other chapters, as it covers a topic that is currently being studied very actively, and this on-going research will likely continue for quite some time. We thus do not have the final word, but instead try to provide some balanced guidance to people who are interested in this subject, in particular those who may want to contribute to it in the future. An unintended consequence of this is the chapter is considerably longer than (most) other chapters.

\begin{table*} [htb]
\centering 
\begin{tabular}{| c | c | c |} 
\hline
Material & Filling factors & References \\
\hline \hline 
\multirow{2}{*}{~Monolayer GaAs~} & ~\multirow{2}{*}{~$5/2, 7/2$~} & 
\cite{Willett1987, PanXia1999} \\
~& ~ & \cite{7/2-2002, Liu-7/2} \\
\hline 
~Bilayer GaAs~ & $1/2$ & ~\cite{331-exp-1, 331-exp-2}  \\
\hline 
~Monolayer ZnO~ & $3/2, 5/2, 7/2, 9/2$ & \cite{Falson2015, Falson2018} \\
\hline 
~Monolayer WSe$_2$~ &  $3/2$ & \cite{Shi-WSe2} \\
\hline 
~Monolayer graphene~ & ~$-1/2$, $1/2$ & \cite{Zibrov2018} \\
\hline
\multirow{3}{*}{~Bilayer graphene~} & ~\multirow{3}{*}{~$-5/2, -1/2, 3/2, 5/2, 7/2$~} & \cite{Ki-graphene, Kim-graphene} \\ 
~& ~ & \cite{Zibrov-bilayer, Li2017} \\
~& ~ & \cite{PRX-graphene2022} \\
\hline 
\end{tabular}
\caption{Summary of observed half-integer FQH states in different materials. Note that additional half-integer FQH states have been observed in higher Landau levels in monolayer graphene~\cite{221_graphene_exp}, which are not listed here.}
\label{tab:half}
\end{table*}

\section{Theoretical background} 
\label{sec:theory}

When a two-dimensional electron gas is placed under a strong perpendicular magnetic field, the electron kinetic energy is quantized into LLs. Consequently, the kinetic energy of electron is quenched, and the energetics is dominated by the strong correlation between electrons~\cite{Yang-Girvin-book}. A minimal Hamiltonian for the FQH regime is \cite{Laughlin1983}:
\begin{eqnarray} \label{eq:H}
H=\frac{e^2}{\epsilon \ell_0} \mathcal{P}_{\text{SLL}} \sum_{i<j} \frac{1}{\vert z_i-z_j \vert} \mathcal{P}_{\text{SLL}}
\end{eqnarray}
where $\mathcal{P}_{\text{SLL}}$ projects the repulsive Coulomb interaction between electrons into a single LL, $z_j=x_j-iy_j$ denotes the complex coordinates of electrons in the $xy$-plane (we assume the magnetic field points upward), $e$ is the electron charge, and $\epsilon$ is the dielectric constant of the host semiconductor. Eq.~\eqref{eq:H} assumes an infinitely thin plane and an infinite LL splitting (no LL mixing).  The former can be easily fixed by including a finite extent of the electron wave function along the perpendicular direction (often referred to as sub-band wave function) when calculating the effective 2D Coulomb potential, while the latter can be corrected by modifying the form of the interaction  \cite{Sarma1997,Jain_book}.  Furthermore, Eq.~\eqref{eq:H} assumes no electron spin.  If we include the spin degree of freedom we must also include a Zeeman term in Eq.~\eqref{eq:H}. For a detailed discussion on different energy scales in FQH states in GaAs heterostructures, and the corresponding justification for Eq.~\eqref{eq:H}, interested readers may refer to~\cite{Papic-review}.

Eq.~\eqref{eq:H} contains only a single interaction term. As a result, standard perturbation theory becomes inapplicable in studying FQH physics, which is solely determined  by $\nu$ (once the interaction is fixed). The seminal work by Jain introduced the theory of composite fermions which is particularly intuitive in revealing how $\nu$ controls the physics~\cite{Jain-CF-theory}.

\subsection{Theory of composite fermions}

\begin{figure} [htb]
\centering
\includegraphics[width=3.3in]{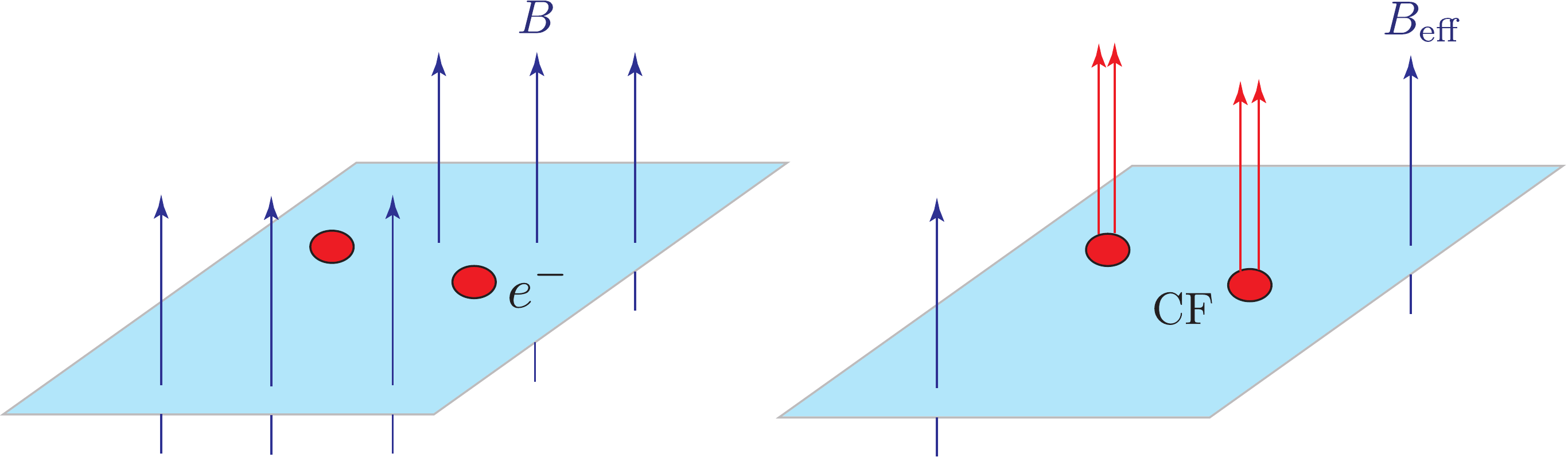}
\caption{Illustration of composite fermions (CFs). Left panel: Electrons in an external magnetic field $B$. Right panel: Composite fermions formed by attaching two magnetic flux quanta to each electron. The effective magnetic field experienced by the composite fermions is $B_{\rm eff}$, which is different from $B$. The number of arrows indicates the number of magnetic flux quanta in the system.}
\label{fig:CF}
\end{figure}

In a first approximation, a composite fermion is formed by attaching an even number $2p$ of magnetic flux quanta to an electron~\cite{Jain_book}. Here, we only focus on the situation in which each composite fermion carries two magnetic flux quanta ($p=1$). The number density of electrons in the two-dimensional electron gas satisfies $\rho=\nu B/\phi_0$, where $\phi_0=h/e$ is the magnetic flux quantum. Due to flux attachment, the composite fermions experience a reduced effective magnetic field $B_{\text{eff}}$. This is illustrated in Fig.~\ref{fig:CF}. Under the mean field approximation, 
\begin{eqnarray} \label{eq:B-eff}
B_{\text{eff}}
=B-2\rho\phi_0
=B(1-2\nu).
\end{eqnarray}
Since $B_{\text{eff}}$ and $B$ are different but the number density of electrons equals that of the composite fermions, the effective filling factor of composite fermions differ from the filling factor of electrons. Then by analogy to the integer quantum Hall effect, the integer quantum Hall state of composite fermions filling $n_{\text{CF}}$ effective LLs (or the so-called $\Lambda$ levels) corresponds to the FQH state of electrons at $\nu=n_{\text{CF}}/(2n_{\text{CF}}\pm 1)$. Here, the sign $\pm$ denotes the direction of the residual magnetic field experienced by the composite fermions. If it is parallel (anti-parallel) to the original field $B$, then the sign is $+ (-)$.

At half filling of the lowest LL, $\nu=1/2$, Eq.~\eqref{eq:B-eff} predicts that the composite fermions experience a zero average magnetic field. This supports the absence of a FQH state at $\nu=1/2$ in monolayer GaAs heterostructures. On the other hand, the ground state of electrons in a half-filled LL is a composite fermion Fermi liquid with a well-defined Fermi surface~\cite{HLR_theory, Kalmeyer1992}. Both the existence of composite fermions and the formation of a composite fermion Fermi surface at $\nu=1/2$ have received strong experimental support~\cite{Goldman-1/2, Hossain2020, Kamburov-exp, Kang_geo-res, Smet-1/2, SAW-half, Willett-half1999}.

A possible starting point for studying FQH states at $\nu>1$ is to simply focus on the partially filled LL(s). The above would then suggest that at $\nu=5/2=2+1/2$ there would exist a gapless state and no FQH. But then how can we understand the presence of an energy gap to support the observed Hall plateau at $\nu=5/2$? Previous approaches such as the hierarchical construction~\cite{Haldane_hierarchy, Halperin_hierarchy} cannot directly explain gapped FQH states at even-denominator filling factors.

Wave function construction is central to the composite fermion ansatz and can be used to make progress. The microscopic flux attachment process defining the composite fermion wave function at total magnetic field $B$ from an effective field at $B_{\text{eff}}$ is given by \cite{Jain-CF-theory}:
\begin{eqnarray} \label{eq:psi-cf}
\Psi^{B}=J^{2p}\psi^{B_{\text{eff}}}
\end{eqnarray}
where $\Psi^B$ is an ansatz state of Eq.~\eqref{eq:H}, $J^{2p}$ is a Jastrow factor attaching an even number of vortices ($2p$) to each electron, and $\psi^{B_{\text{eff}}}$ is a fermionic wave function at an effective field $B_{\text{eff}}$ that results from projection into a single LL \cite{Jain1997}. In the symmetric gauge we have, for $N$ particles:
\begin{eqnarray} \label{eq:jastro}
J=\prod_{i<j}^N(z_i-z_j)
\end{eqnarray} 
where we see that $J$ vanishes if two particles occupy the same location. $J$ implements vortex attachment \cite{Read1989,Read1994,Read1996} which can be approximated by flux attachment in effective theories.  Here we have used the terms vortex and flux attachment interchangeably with the understanding that they are not exactly equivalent \cite{Murthy2003a, Jain_book}.

$\Psi^{B}$ describes a large class of possible ansatz ground states because we are free to choose $\psi^{B_{\text{eff}}}$.  In the simplest setting we assume that $\psi^{B_{\text{eff}}}$ describes weakly interacting fermions.  For example, Eq.~\eqref{eq:psi-cf} reduces to the Laughlin wave function at $\nu=1/(2p+1)$ \cite{Laughlin} if we set $\psi^{B_{\text{eff}}}$ to be an integer filled state of composite fermions at effective filling of $n_{\text{CF}}=1$ \cite{Jain-CF-theory,Jain_book}.  We therefore see that the Laughlin states form a subset of states within the composite fermion wave function formalism.  

A large class of states at half filling are also captured by $\Psi^{B}$.  The composite fermion Fermi liquid \cite{HLR_theory} arises when we assume a filled Fermi sea at zero effective field ($n_{\text{CF}}\rightarrow\infty$) \cite{Rezayi1994a} for $\psi^{B_{\text{eff}}}$. Here we have a gapless state that accurately captures the essential physics of the half filled lowest LL \cite{Jain_book}.  But in the following we discuss gapped examples of $\psi^{B_{\text{eff}}}$ relevant for even filling, specifically paired states, that yield energetically competitive ansatz wave functions for half filling of the second LL. 

\subsection{Paired quantum Hall states}

\begin{figure} [htb]
\centering
\includegraphics[width=2.75in]{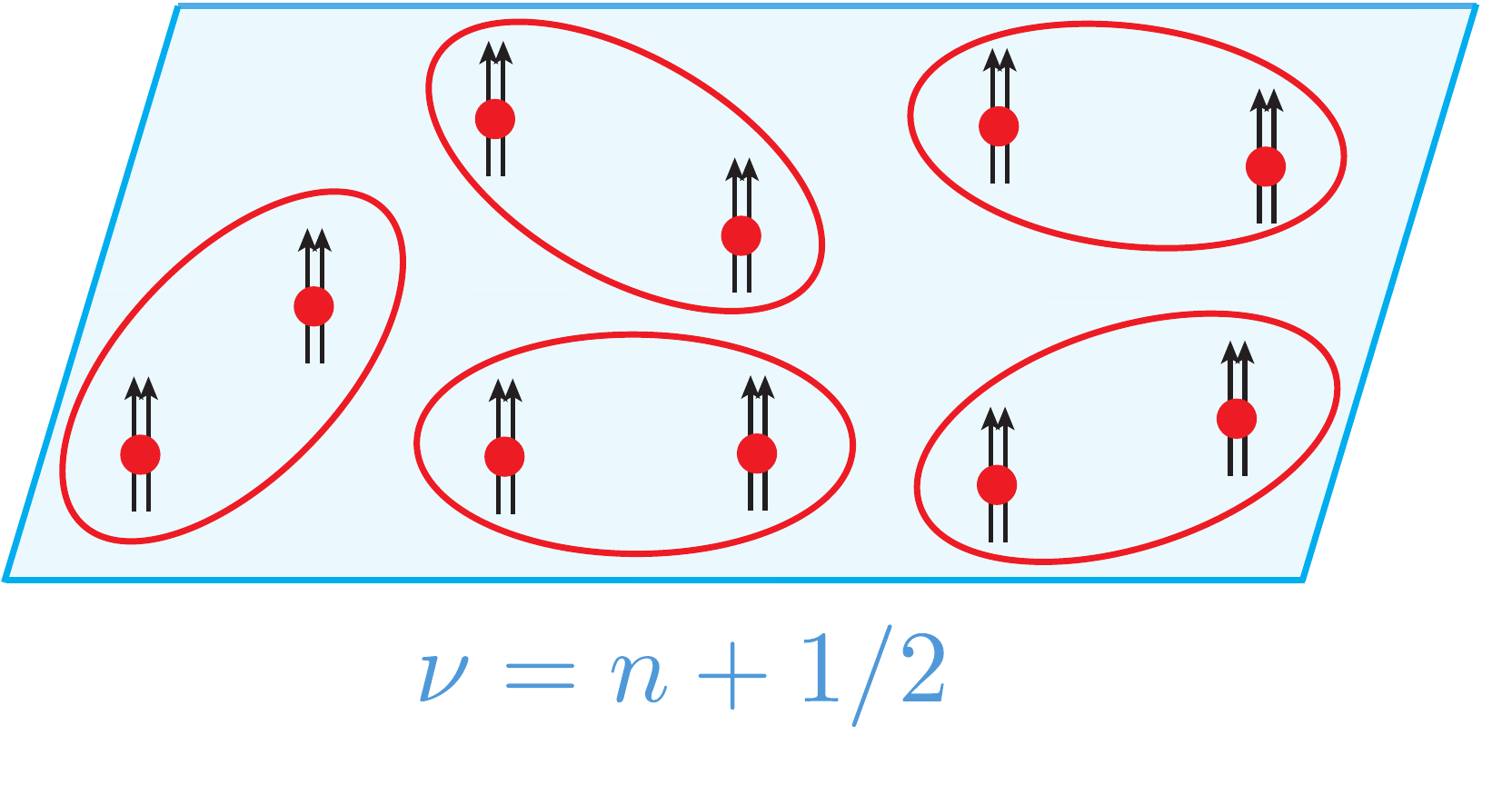}
\caption{Pairing of composite fermions in the $\nu=5/2$ and other half-integer FQH states. The red dots are electrons. Each of them is attached to two magnetic flux quanta marked by black arrows and form a composite fermion. These composite fermions form Cooper pairs (grouped by the red ovals in the figure), that gives rise to a gapped phase at half-integer filling factors.}
\label{fig:CF-pair}
\end{figure}

The even denominator in the filling factor indicates that the $5/2$ state may originate from bosonic entities formed by electrons or composite fermions. Different from the original description with strongly correlated electrons, the effective interaction between composite fermions is much weaker. Similar to the case of $\nu=1/2$, a composite fermion Fermi liquid is formed when the system has filling factor $5/2$. However, the effective interaction between composite fermions becomes weakly attractive at $\nu=5/2$ and leads to a Cooper instability~\cite{Scarola_nature}. It is believed that the pairing between composite fermions gives rise to the gapped FQH state at $\nu=5/2$ and other half-integer filling factors. This idea is illustrated in Fig.~\ref{fig:CF-pair}.  This motivates ansatz states where $\psi^{B_{\text{eff}}}$ is an eigenstate of an effective pairing Hamiltonian \cite{Read-Green}. 

The pairing Hamiltonian of Bardeen-Cooper-Schrieffer (BCS) takes the form:
\begin{align} \label{eq:H-BCS}
\nonumber
H_{\text{BCS}}
=&\sum_{\bm{k}}
\left[\left(\frac{\hbar^2 k^2}{2m^*}-\mu\right)
c_{\bm{k}}^\dagger c_{\bm{k}}\right.
\\
&+\left.\frac{1}{2}\left(\Delta^*_{\bm{k}} c_{-\bm{k}}c_{\bm{k}}
+\Delta_{\bm{k}}c_{\bm{k}}^\dagger c_{-\bm{k}}^\dagger\right)\right].
\end{align}
Here, $c_{\bm{k}}$ and $c_{\bm{k}}^\dagger$ are the second quantized operators for composite fermions, with the index $\bm{k}$ labeling the momentum. We assume the composite fermions are spin-polarized in this section and have thus suppressed the spin index. The symbols $m^*$ and $\mu$ label the effective mass and chemical potential of composite fermions, respectively. Importantly, both $m^*$ and $\mu$  are non-perturbative and  due solely to the Coulomb interaction between electrons.

Knowing that composite fermions can form Cooper pairs does not completely determine the nature of the $5/2$ state. Different pairing channels with different gap functions $\Delta_{\bm{k}}$ lead to different possible phases \cite{Read-Green}. Which one or what combination of these states are actually realized in a realistic experimental sample may depend sensitively on the microscopic details and experimental conditions. Notice that we  use the terminologies, ``states" and ``topological orders" (sometimes abbreviated as orders) interchangeably in the following discussion.

\subsection{Pfaffian (Moore-Read) state} 

The Pfaffian state (also known as the Moore-Read state) has the following wave function~\cite{MR1991}:
\begin{eqnarray} \label{eq:MR}
\Psi_{\text{Pf}}(\left\{z_i\right\})
=
J^2 \text{Pf}\left(\frac{1}{z_i-z_j}\right)
\exp{\left(-\sum_i\frac{|z_i|^2}{4\ell_0^2}\right)}.
\end{eqnarray}
The attachment of two flux quanta to each electron is implemented by the Jastrow factor, which also fixes the filling factor to be $\nu=1/2$. The Gaussian factor is ubiquitous in every quantum Hall wave function, which originates from the eigenfunctions of LLs in the symmetric gauge~\cite{Prange-book}.  This factor will be dropped in the following discussion. Finally, the Pfaffian factor is defined as
\begin{eqnarray}
\text{Pf}\left(\frac{1}{z_i-z_j}\right)
=\mathcal{A}\left(\frac{1}{z_1-z_2}\frac{1}{z_3-z_4}\cdots\right),\label{eq:Pf-def}
\end{eqnarray}
where $\mathcal{A}$ denotes antisymmetrization over the set of coordinates $z_i$. This ``explains" the name of the Pfaffian state. 

The Pfaffian wave function has several important features.   The right hand side of Eq.~\eqref{eq:Pf-def} can be understood as a real-space BCS paired state at zero effective magnetic field~\cite{Read-Green, Schrieffer2018}.  $\Psi_{\text{Pf}}$ is therefore gapped owing to the composite fermion BCS gap.  $\Psi_{\text{Pf}}$ is also an exact zero energy ground state of an idealized three-body Hamiltonian~\cite{Greiter1992}, which will be further discussed in Sec.~\ref{sec:numerics}. 

Eq.~\eqref{eq:MR} is an ansatz state for the incompressible (gapped) state of electrons half filling of the lowest LL. To describe the 5/2 state, we assume the electrons completely fill the lowest two LLs (one with spin up and another with spin down polarizations), and the Pfaffian wave function describes the fully polarized electrons in the half filled second LL only~\cite{Greiter1992, MR1991,Read-Green}. Note that wave functions for FQH states in higher LLs can be generated systematically by acting LL raising operators on a wave function defined in the lowest LL~\cite{Yang-Girvin-book}. For this reason we can always represent a wave function in any LL by its lowest LL counterpart, which is what we do throughout this chapter unless noted otherwise.

\subsubsection{Conformal field theory description}

Historically, the wave function in Eq.~\eqref{eq:MR} was introduced in~\cite{MR1991} via conformal field theory (CFT). CFT has since become a powerful tool in studying quantum Hall physics~\cite{Hansson-CFT}. For details on CFT, readers may refer to the textbook~\cite{CFT-book}. Here, we will only outline the basic ideas and state the necessary results directly.

For the Pfaffian state, the CFT description actually involves two different CFTs. The first part is a compactified holomorphic boson $\phi(z)$, which is responsible for the correct filling factor at $\nu=1/2$ for the half-filled LL. The second part is the holomorphic Ising CFT which has three different primary fields as summarized in Table~\ref{tab:Ising-CFT}. In particular, there is a holomorphic Majorana fermion field $\psi(z)$ which has the scaling dimension $1/2$. The electron operator is defined as
\begin{eqnarray}
\hat{\Psi}_e(z_j)=\psi(z_j)\exp{[2i\phi(z_j)]},
\end{eqnarray}
which has scaling dimension $3/2$ indicating that it indeed represents a fermion. The wave function in Eq.~\eqref{eq:MR} can now be interpreted as the correlation function between electron operators:
\begin{eqnarray}
\Psi_{\text{Pf}}(\left\{z_j\right\})
=\prod_{j=1}^N
\Big\langle
\hat{\Psi}_e(z_j)\hat{O}_{\text{bg}}(z_j)
\Big\rangle.
\end{eqnarray} 
In the above equation, the background charge operator $\hat{O}_{\text{bg}}(z_j)$ ensures that the correlation function does not vanish. It is also responsible for reproducing the correct Gaussian factor in the wave function.

\begin{table*} [htb]
\begin{subtable}[c]{0.5\textwidth}
\centering
\begin{tabular}{|c|c|}
\hline
~Primary field~ & ~Scaling dimension~ \\
\hline\hline
 $I$ & $0$   \\
\hline 
 $\psi$ & $1/2$ \\
\hline
$\sigma$ & $1/16$ \\
\hline
\end{tabular}
\end{subtable}
\begin{subtable}[c]{0.5\textwidth}
\centering
\begin{tabular}{|c||c|c|c|}
\hline
~$\times$~ & ~$I$~ & ~$\psi$~ & ~$\sigma$~ \\
\hline\hline
~$I$~ & ~$I$~ & ~$\psi$~ & ~$\sigma$~  \\
\hline 
~$\psi$~ & ~$\psi$~ & ~$I$~ & ~$\sigma$~ \\
\hline
~$\sigma$~ & ~$\sigma$~ & ~$\sigma$~ & ~$\psi+I$~ \\
\hline
\end{tabular}
\end{subtable}
\caption{Primary fields in the holomorphic Ising CFT (left table) and their fusion rules (right table).}
\label{tab:Ising-CFT}
\end{table*}

After defining the suitable electron operator in the CFT approach, all possible quasiparticles in the quantum Hall state can be deduced systematically. For example, consider the quasiparticle described by the CFT operator 
\begin{eqnarray}
\hat{\Psi}_{\text{qp}}(z)
=\sigma(z) e^{i\alpha\phi(z)},
\end{eqnarray}
where $\alpha$ is a constant that needs to be determined. The operator product expansion between the quasiparticle and the electron operators must be single-valued (or without any branch cut). This condition leads to $\alpha=n'+1/2$, with $n'$ being an arbitrary integer. In particular, $n'=0, -1$ corresponds to quasiparticles with charges $\pm e/4$. The corresponding CFT operator has scaling dimension $1/8$, which is the most relevant operator among all possible operators for quasiparticles in the Pfaffian state. Interestingly, the fusion rule $\sigma\times\sigma=\psi+I$ indicates that these quasiparticles are non-Abelian anyons. This exciting feature has motivated many of the ongoing studies of $5/2$ state, in the hope of finding its potential applications in topological quantum computation~\cite{Kitaev2003, Das-Sarma}. 

\subsubsection{Edge theory}

FQH states that are gapped in the bulk must nonetheless become gapless at the edges \cite{Halperin1982}. Effective low-energy theories have been used to model gapless edge states~\cite{Wen_book}. Also, the bulk-edge correspondence suggests that different bulk topological orders usually have correspondingly different edge structures~\cite{Cano2014, Dubail_Read, Qi2012, Swingle2012, Yan2019}. The knowledge of edge physics can provide invaluable insight into the nature of the underlying quantum Hall states. Hence, various types of experiments have been proposed and performed to probe the edge structure of different quantum Hall states. For a review of these edge probes, see~\cite{Heiblum-Feldman}.

\begin{figure*} [htb]
\centering
\includegraphics[width=6.0in]{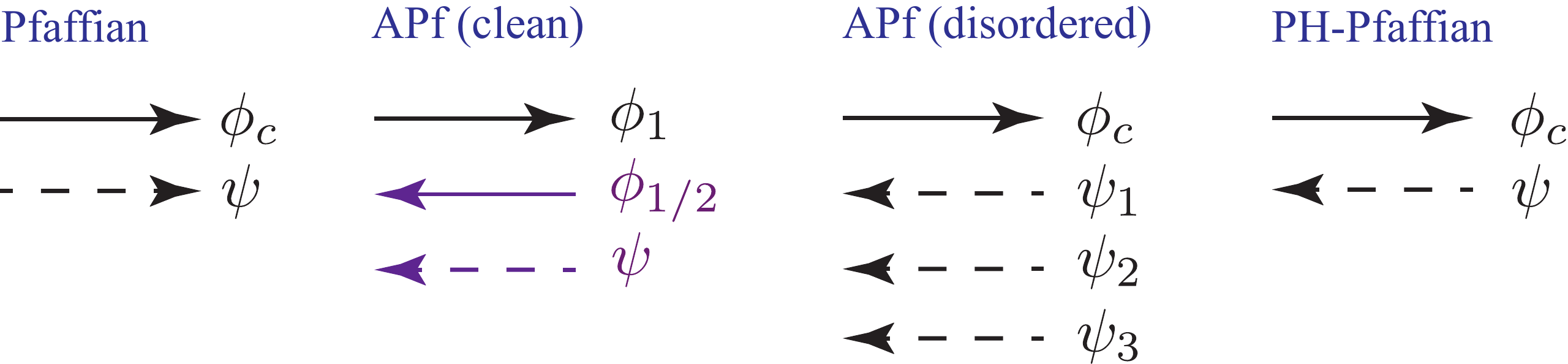}
\caption{Edge structures of the Pfaffian, anti-Pfaffian (abbreviated as APf in the figure), and PH-Pfaffian states. Here, solid lines and dashed lines denote Bose modes and Majorana fermion modes, respectively. The chirality of the overall charge mode $\phi_c$ is defined as the downstream direction (pointing to the right). In the anti-Pfaffian state, the description of its edge structure depends on disorder. In the case of clean edge, there are two Bose charge modes $\phi_1$ (integer mode) and  $\phi_{1/2}$ (fractional mode). These two modes can be coupled by disorder. For sufficiently strong disorder, the edge is driven to the disorder-dominated fixed point and being described by an overall charge mode and three upstream Majorana fermion modes. Note that we have neglected the two additional integer Bose modes due to the two lower LLs which are present in the actual edge of the $5/2$ state.}
\label{fig:edge}
\end{figure*}

For the Pfaffian state, its edge structure can be deduced from the aforementioned CFT approach. Similar to the Laughlin state, the Pfaffian edge possesses a chiral Bose mode $\phi_c$. The subscript $c$ indicates that it is a charge mode. Its chirality, which corresponds to the downstream direction of propagation, is fixed by the direction of the external magnetic field. In addition, the Pfaffian edge has a chiral Majorana fermion mode $\psi$ that is copropagating with the Bose mode~\cite{Milovanovic1996, Wan-Pf2006, Wen-Pf1993}. As a result, the Pfaffian edge is described by the Lagrangian density,
\begin{eqnarray}
\mathcal{L}_{\text{Pf}}
=-\frac{2}{4\pi}\partial_x\phi_c (\partial_t+v_c\partial_x)\phi_c
+i\psi(\partial_t+v_n\partial_x)\psi.
\end{eqnarray}
The symbols $v_c$ and $v_n$ denote the speeds of the charge Bose mode and neutral Majorana fermion mode, respectively. Notice that a chiral Bose field and a chiral Majorana fermion field are present in both the bulk and the edge of the Pfaffian state. This feature is a manifestation of the bulk edge correspondence. In numerical simulations, $v_c$ and $v_n$ were found to satisfy $v_c\gg v_n$ due to Coulomb interaction~\cite{Yang-Pf2008} --- see Sec.~\ref{sec:numerics}. The charge density of the edge is completely governed by the Bose mode, given by $\rho(x)=-e\partial_x\phi_c/2\pi$. Its corresponding edge channel contributes to an electrical Hall conductance $\sigma_{xy}=e^2/2h$. At $\nu=5/2$, there are actually two more edge channels originating from the two completely filled LLs. Both of them are described by chiral Bose modes different from $\phi_c$. Each of these integer channels contributes $\sigma_{xy}=e^2/h$, so the electrical Hall conductance $5e^2/2h$ at $\nu=5/2$ follows. Unless otherwise specified, we will ignore the two integer channels in this section. The edge structure for the Pfaffian state is shown in Fig.~\ref{fig:edge}.

\subsubsection{Weak pairing of composite fermions}

As previously mentioned, the FQH state at $\nu=5/2$ is believed to originate from the pairing between composite fermions. The Pfaffian state in particular postulates a weak pairing phase [i.e., $\mu>0$ in Eq.~\eqref{eq:H-BCS}] of composite fermions in the $p$-wave angular momentum channel~\cite{Read-Green}. The corresponding gap function in Eq.~\eqref{eq:H-BCS} takes the form $\Delta_{\bm{k}}=\Delta_0(k_x+ik_y)$. From this perspective, the Pfaffian factor in the wave function can be interpreted as the BCS wave function in position space representation. Furthermore, the Chern number of this weak pairing phase was evaluated to be $\mathcal{C}=1$; an explicit calculation can be found in the Appendix A of~\cite{Ken-1/4}. This indicates that the Pfaffian state is topologically nontrivial. It is reflected by the existence of a chiral Majorana fermion mode on the edge, which is also consistent with the breaking of time reversal symmetry in the  $p$-wave BCS pairing channel.

\subsection{Anti-Pfaffian state}

When the effect of LL mixing is negligible, the particle-hole transformation is an exact symmetry in a half-filled LL. Thus, the particle-hole conjugate of the Pfaffian state provides another viable candidate to describe the $5/2$ state. This motivated the introduction of the anti-Pfaffian state in~\cite{Apf_Levin2007} and~\cite{Apf_Lee2007}. In principle, the wave function of the anti-Pfaffian state can be obtained systematically from particle-hole conjugating the Pfaffian wave function in Eq.~\eqref{eq:MR}~\cite{Apf_Lee2007}. In fact, as discussed in Sec.~\ref{sec:numerics}, since most numerical studies consider the limit where particle-hole symmetry is exact, both the Pfaffian and anti-Pfaffian states are trivially related --- they are particle-hole conjugate states to one another. Nevertheless, more recent numerical work has undertaken studies of the anti-Pfaffian state under various conditions. 

Using the parton theory introduced in~\cite{Jain-221}, the $\bar{2}\bar{2}111$ parton state was proposed to describe the $5/2$ state~\cite{Balram-APf2018}. It was discovered that the $\bar{2}\bar{2}111$ parton state describes the same phase, and has a high numerical overlap with the wave function of the anti-Pfaffian state~\cite{Balram-APf2018}. Meanwhile, the parton state is much easier to use in numerics.

The anti-Pfaffian state has a more complicated edge structure than the Pfaffian. First, there is an integer Bose mode circulating on the edge in the downstream direction. Due to the particle-hole transformation, the chiralities of the Bose mode and the Majorana fermion mode on the original Pfaffian edge are reversed. As a result, there are three different edge modes in the anti-Pfaffian state as illustrated in Fig.~\ref{fig:edge}. In a real experimental sample, disorder exists and couples the edge modes. When this coupling is sufficiently strong, the anti-Pfaffian edge is driven to the disorder-dominated phase, which consists of a single downstream charge Bose mode and three upstream neutral Majorana fermion modes~\cite{Apf_Lee2007, Apf_Levin2007}. Furthermore, these Majorana fermion modes exhibit an emergent SO$(3)$ symmetry and have the same velocity $\bar{v}_n$. As a result, the edge can be described by
\begin{eqnarray} \label{eq:Apf-edge}
\begin{aligned}
\mathcal{L}_{\text{Apf}}
=~&-\frac{2}{4\pi}\partial_x\phi_c(\partial_t+v_c\partial_x)\phi_c
\\
&+i\sum_{j=1}^3\psi_j (\partial_t-\bar{v}_n\partial_x)\psi_j.
\end{aligned}
\end{eqnarray}
The anti-Pfaffian state also hosts non-Abelian anyons with charges $\pm e/4$. Since the edge is not maximally chiral, the scaling dimensions of CFT operators for quasiparticles are non-universal. The $e/4$ anyon has a minimum scaling dimension of $1/4$, which is achieved in the aforementioned disorder-dominated phase. The scaling dimension is larger than the value of $1/8$ in the Pfaffian state. This difference turns out to be important when one considers tunneling experiments discussed in Sec.~\ref{sec:old-exp}.

In the weak pairing picture, the anti-Pfaffian state is interpreted as the superconducting phase arising from $f$-wave pairing between composite fermions. The gap function in Eq.~\eqref{eq:H-BCS} takes the form $\Delta_{\bm{k}}=\Delta_0(k_x-ik_y)^3$. Note that the corresponding Bogoliubov-de Gennes (BdG) Hamiltonian has Chern number $\mathcal{C}=-3$. This explains the difference between the numbers of chiral Majorana fermion edge modes in Pfaffian and anti-Pfaffian states.

\subsection{PH-Pfaffian state}

Recently, the particle-hole Pfaffian (PH-Pfaffian) state has attracted a considerable amount of attention as a candidate $5/2$ state in GaAs heterostructures. It was originally introduced in the theory of Dirac composite fermions, which is a theory that preserves particle-hole symmetry explicitly in the half-filled LL~\cite{Son2015, Son2016}. In particular, the PH-Pfaffian state can be viewed as arising from $s$-wave pairing between Dirac composite fermions. This translates into a $p$-wave pairing (with the opposite chirality as the Pfaffian) between ordinary composite fermions, in which the corresponding BdG Hamiltonian has Chern number $\mathcal{C}=-1$. Different from the Pfaffian and anti-Pfaffian states, the PH-Pfaffian state is particle-hole symmetric. It is worth noting that a similar state known as the time-reversal symmetry preserving Pfaffian ($\mathcal{T}$-Pfaffian) was introduced to describe the $\mathcal{T}$-invariant surface state of topological insulators~\cite{x18b-Bonderson, x18a-Ashvin}. Nevertheless, the realization of PH-Pfaffian order does not require the system itself to preserve particle-hole symmetry~\cite{Zucker2016} --- after all, LL mixing and other effects present in real experimental systems explicitly break particle-hole symmetry.

The edge structure of the PH-Pfaffian state can be deduced from its particle hole symmetry. After a particle hole transformation, the resulting edge theory has central charge $1-c$, with $c$ being the central charge of the original edge theory. In order for the edge to be particle-hole symmetric, $c$ needs to be $1/2$. Furthermore, there must be a downstream Bose mode $\phi_c$, such that its corresponding edge channel leads to the correct electrical Hall conductance of $\sigma_{xy}=e^2/2h$. Therefore, the remaining edge mode must be an upstream Majorana fermion mode $\psi$. Hence, the PH-Pfaffian edge is described by the Lagrangian density,
\begin{align}
\nonumber
\mathcal{L}_{\text{PH-Pf}}
=~&-\frac{2}{4\pi}\partial_x\phi_c(\partial_t+v_c\partial_x)\phi_c
\\
&+i\psi(\partial_t-v_n\partial_x)\psi.
\end{align}
The edge structure is shown in Fig.~\ref{fig:edge}.

After deducing the edge structure, it is straightforward to formulate the CFT description of the bulk. The electron operator with the appropriate chirality and scaling dimension is $\hat{\Psi}_e(z)=\psi(\bar{z})\exp{[2i\phi(z)]}$. Here, $\bar{z}$ denotes the complex conjugation of $z$. Physically, the antiholomorphicity of $\psi$ originates from its upstream chirality at the edge. By evaluating the correlation function between electron operators, a possible wave function for the PH-Pfaffian state was written down explicitly in~\cite{Zucker2016}:
\begin{eqnarray}
\label{eq:PHPf}
\Psi_{\text{PH-Pf}}\left(\left\{z_i\right\}\right)
=J^2\text{Pf}\left(\frac{1}{\bar{z}_i-\bar{z}_j}\right).
\end{eqnarray}
To be studied in numerics, one needs to project the wave function into a single LL, which turns out to be subtle. An alternative wave function can be obtained based on the argument of negative flux attachment~\cite{Jolicoeur2007}. However, there is limited numerical support for either wave function to describe a gapped phase after the lowest LL projection. This issue will be discussed more in Sec.~\ref{sec:numerics}.

\subsection{Unified description: Kitaev's 16-fold way}

Based on different approaches such as trial wave functions, CFT, parton theory, and particle-hole conjugation, other possible candidates for describing FQH states at half-integer filling factors were introduced. They are listed in Table~\ref{tab:summary}. Some of them may be irrelevant to the $5/2$ state in monolayer GaAs quantum wells, but they may describe different half-integer states in other materials listed in Table~\ref{tab:half}. For example, the spin-unpolarized version of Halperin-331 state is believed to describe the FQH state at $\nu=1/2$ in bilayer systems and wide quantum wells. It has been studied widely numerically~\cite{331-num-2, 331-num-3, Peterson-Papic-DS, Scarola-PRB2010, SJ2001, 331-num-1}. Also, the 221 parton state may describe half-integer states in graphene systems~\cite{221_graphene_exp, 221_graphene}. Given that these topological orders were introduced via different approaches, it is desirable to have a unified and systematic description.

The possible types of topological orders or phases in the paired quantum Hall state depend on the spin polarization of electrons and the associated composite fermions.  Experimental results suggest that the $5/2$ state observed in GaAs samples at the electron density $\rho\sim (2-3)\times 10^{11}$cm$^{-2}$ are spin polarized (see Sec.~\ref{sec:old-exp}). This conclusion is also supported from numerical work (see Sec.~\ref{sec:numerics}). Furthermore, a recent geometric resonance measurement performed very close to $\nu=5/2$ has provided strong evidence of a fully spin-polarized Fermi sea for composite fermions~\cite{polarized-CF}. 

There are an infinite number of possible channels for pairing composite fermions which would, at first, appear to give an infinitely long list of possible topological orders for describing the $5/2$ state. But the classification of possible theories is constrained. The classification of topological orders is based on the types of anyons they possess~\cite{Wen_book}. When two topological orders have the same set of anyons (i.e., all fractional charges, braiding statistics, and fusion rules are the same), then they are  equivalent. Following the above definition, there are only 16 possible topological orders from pairing spin-polarized composite fermions. This property is known as  Kitaev's 16-fold way~\cite{Kitaev}. Therefore, most of the existing candidates for describing the spin-polarized $5/2$ state are particular members in the 16-fold way~\cite{16-fold}. The original work by Kitaev considered topological orders with neutral excitations only, however, charge excitations exist in the $5/2$ state. It was shown that all of these topological orders can host charge $\pm e/4$ anyons. The Chern number $\mathcal{C}$ of the topological order carries important information, which will be discussed below. Based on the 16-fold way description, various experimental signatures for different topological orders were predicted and summarized in Table~\ref{tab:summary}.

\begin{table*}[t]
\centering
\begin{tabular}{| c | c | c | c | c | c | c |}
\hline
~$\mathcal{C}$~ & ~Name~ & ~References~ & ~$g_{e/4}$~ & ~$\kappa_H$~ & ~Even-odd effect?~ & ~Shift~ 
\\ \hline\hline
~$0$~ & ~$K=8$ & ~(i, ii)~ &  ~$\bf{1/8}$~ & ~$3$~ & ~No~ & 3
\\  \hline
~$1$~ & ~Pfaffian & ~(iii, iv)~ &  ~$\bf{1/4}$~ & ~$3.5$~ & ~Yes~ & 3
\\  \hline
~$-1$~ & ~PH-Pfaffian & ~(v, vi)~ &  ~$\bf{1/4}$~ & ~$2.5$~ & ~Yes~ & 1
\\  \hline
~$2$~ & ~$331$ & ~(i)~ & ~$\bf{3/8}$~ & ~$4$~ & ~Maybe~ & 5
\\  \hline
~$-2$~ & ~$113$ & ~(vii, viii)~ & ~$\bf{3/8}$~ & ~$2$~ & ~Maybe~ & -1
\\  \hline
~$3$~ & ~221 parton/ SU(2)$_2$ & ~(ix, x)~ & ~$\bf{1/2}$~ & ~$4.5$~ & ~Yes~ & 5
\\  \hline
~$-3$~ & ~Anti-Pfaffian/ $\bar{2}\bar{2}111$ parton & ~(ix, xi, xii, xiii)~ & ~$\bf{1/2}$~ & ~$1.5$~ & ~Yes~ & $-1$
\\  \hline
~$4$~ &  ~$\slash$ & ~$\slash$ & ~$5/8$~ & ~$5$~ & ~Maybe~ & $\slash$
\\  \hline
~$-4$~ & ~Anti-331 & ~(xiv)~  & ~$5/8$~ & ~$1$~ & ~Maybe~ & $\slash$
\\ \hline
 ~$5$~ & ~$\slash$ & ~$\slash$ & ~$3/4$~ & ~$5.5$~& ~Yes~ & 7
\\ \hline
~$-5$~ & ~Anti-SU(2)$_2$ & ~(xiv)~ & ~$3/4$~ & ~$0.5$~ & ~Yes~ &  -3
\\ \hline
~$6$~ & ~$\slash$ & $\slash$ & ~$7/8$~ & ~$6$~ & ~Maybe~ & $\slash$
\\ \hline
~$-6$~ & ~$\slash$ & $\slash$ & ~$7/8$~ & ~$0$~ & ~Maybe~ & $\slash$
\\ \hline
~$7$~ & ~$\slash$ & $\slash$ & ~$1$~ & ~$6.5$~ & ~Yes~ & 9
\\ \hline
~$-7$~ & ~$\slash$ & $\slash$ & ~$1$~ & ~$-0.5$~ & ~Yes~ & -5
\\ \hline
~$8$~ & ~$\slash$ & $\slash$ & ~$9/8$~ & ~$7$~ & ~Maybe~ & $\slash$
\\ \hline
\end{tabular}
\caption{Possible topological orders in the 16-fold way for the spin-polarized $5/2$ state, and their predicted experimental signatures and topological shifts. Here, the possibilities of edge reconstruction and Majorana gapping are ignored. The first column shows the Chern numbers of the topological orders. Topological orders with odd (even) $\mathcal{C}$ are non-Abelian (Abelian). On the edges of these topological orders, there are $|\mathcal{C}|$ chiral Majorana fermion modes. The orders with $\mathcal{C}<0$ have upstream neutral modes. Since a pair of Majorana fermion modes can be combined into a neutral Bose mode, there is no unpaired Majorana fermion mode at even $\mathcal{C}$. In the second and third columns, we provide the usual names of the topological orders in the existing literature if any (see the end of this caption for the key of the references). The smallest possible charge anyon has the charge $e/4$. When the edge is in the disorder-dominated phase, the fourth column gives the tunneling exponents $g_{e/4}=2\Delta_{e/4}$ for the $e/4$ anyons. These values are boldfaced when the $e/4$ anyon is the most relevant quasiparticle. The fifth column lists the thermal Hall conductance in units of $\kappa_0=\pi^2 k_B^2 T/3h$. They are half-integers (integers) for non-Abelian orders (Abelian orders). All non-Abelian orders should demonstrate the even-odd effect in a Fabry-P\'{e}rot interferometer. Abelian orders (except the $K=8$ state) may also show the same effect if they possess flavor symmetry. These are listed in the sixth column. The last column shows the topological shifts for different topological orders, which are important for numerical studies performed on the spherical geometry. Note that the shifts for Abelian orders here are for the spin-polarized states. For the spin-unpolarized Halperin-331 state, the shift is $3$ but not $5$~\cite{Wen-Pf1993}. For Abelian orders with $|\mathcal{C}|\geq 4$, the shifts are unpublished. A brief overview of different experimental techniques are given in Secs.~\ref{sec:old-exp}-\ref{sec:future}. For a detailed discussion, see~\cite{16-fold}. The labels for the references in which the topological orders were introduced are (i)~\cite{Halperin331}; (ii)~\cite{Wen_Zee}; (iii)~\cite{MR1991}; (iv)~ \cite{Greiter1991}; (v)~\cite{Son2015}; (vi)~\cite{Zucker2016}; (vii)~\cite{Guang2014}; (viii)~\cite{Guang2015}; (ix)~\cite{Jain-221}; (x)~\cite{Jain-221-2}; (xi)~\cite{Apf_Levin2007}; (xii)~\cite{Apf_Lee2007}; (xiii)~\cite{Balram-APf2018}; (xiv)~\cite{Guang2013}.}
\label{tab:summary}
\end{table*}

\subsubsection{Properties of anyons}
\label{sec:propanyons}

$\mathcal{C}$ can be identified as the Chern index in the original Kitaev's 16-fold way, such that the properties of anyons in each topological order can be inferred. The eight orders with even $\mathcal{C}$ are Abelian, whereas the eight with odd $\mathcal{C}$ are non-Abelian. In the latter case, quasiparticles with charges $(2m+1)e/4$ (where $m\in\mathbb{Z}$) are non-Abelian anyons. For the former case, there are two different types (or flavors) of Abelian anyons for quasiparticles with charges $(2m+1)e/4$. The scaling dimensions of the electron operator, and the operators for quasiparticles with charges $\pm e/4$ and $\pm e/2$ are~\cite{16-fold}
\begin{eqnarray}
\Delta_e=\frac{3}{2}
~,~
\Delta_{e/2}=\frac{1}{4}
~,~ \mathrm{and}~
\Delta_{e/4}=\frac{|\mathcal{C}|+1}{16}.
\end{eqnarray}
These results are useful in analyzing quasiparticle tunneling conductance experiments. In particular, the low-temperature electrical conductance, in the idealized case, obeys $\sigma\sim T^{2g-2}$, where $g=2\Delta$ is the tunneling exponent~\cite{Wen_book} and $T$ is temperature. The value of $g$ depends on the type of quasiparticles participating in the tunneling process, which can have different scaling dimension $\Delta$ for different candidate states for the $5/2$ FQH state.

\subsubsection{Edge theory}

The Chern number $\mathcal{C}$ also specifies the number and chirality of the Majorana fermion modes on the edge of the topologically ordered phase. Suppose the edge is equilibrated by disorder, then its appropriate description consists of an overall charge mode $\phi_c$ and a collection of Majorana fermion modes $\psi_j$. Since a pair of counter propagating Majorana fermion modes can be gapped out or become localized due to electron tunneling on the edge, we can assume all $\psi_j$ have the same chirality. In addition, these $\psi_j$ exhibit an emergent SO$(|\mathcal{C}|)$ symmetry in the low temperature limit (i.e., the symmetry breaking terms are irrelevant in the renormalization group sense) and have the same velocity $\bar{v}_n$. For more details, see~\cite{Guang2013, Guang2014}. From the above discussion, the edge of a topologically ordered phase with Chern number $\mathcal{C}$ is described by~\cite{Dima-fractional, 16-fold}:
\begin{align}
\nonumber
\mathcal{L}
=&~-\frac{2}{4\pi}\partial_x\phi_c(\partial_t+v_c\partial_x)\phi_c
\\
&+i\sum_{j=1}^{|\mathcal{C}|}\psi_j
\left[\partial_t+\text{sgn}(\mathcal{C})\bar{v}_n\partial_x\right]\psi_j.
\end{align}

Since a pair of $\psi_j$ can be combined into a complex fermion mode and becomes a Bose mode upon bosonization, there are $|\mathcal{C}|/2$ neutral chiral Bose modes when $\mathcal{C}$ is even. Hence, Abelian orders such as the unpolarized Halperin-331 state do not have unpaired Majorana fermion modes on the edge. The existence or absence of unpaired Majorana fermion modes can be probed in the thermal Hall conductance experiments. The edge structures are different for topological orders with $\mathcal{C}$ and $\mathcal{C}+16m$ (where $m\in\mathbb{Z}$), but the 16-fold way implies that they have the same set of anyons. Hence, they are equivalent.

Edge reconstruction and Majorana fermion gapping lead to additional candidates for describing the $5/2$ edge state. These include the Majorana-gapped edge-reconstructed Pfaffian state and the Majorana gapped anti-Pfaffian state~\cite{Overbosch2008}. However, their realization in real GaAs samples is not supported by tunneling conudctuance experiments or upstream noise measurements (see Sec.~\ref{sec:old-exp}).

\subsection{Bulk collective excitations}
\label{sec:collective}

Incompressible FQH states must be gapped at all wavevectors.  It turns out that lowest LL FQH states at filling $n_{\text{CF}}/(2pn_{\text{CF}}\pm1)$ possesses a low-lying intra-LL branch of collective modes that were first identified numerically at 1/3 filling \cite{Haldane1985a}.  These modes are accurately captured by the composite fermion wave functions.  

Quasiparticle-quasihole pairs inserted into  $\psi^{B_{\text{eff}}}$ define composite fermion magnetoexcitons.  Composite fermion magnetoexcitons have a quasiparticle-quasihole spacing proportional to their wavevector. The ensuing collective mode dispersion reveals neutral roton modes at intermediate wavevectors and a well separated quasiparticle-quasihole pair at large wavevectors.  (Alternatively, they can be viewed as a density wave created on top of the otherwise uniform ground state.)  

Composite fermion magnetoexciton wave functions have been systematically compared against unbiased numerical results and have been found to be essentially exact in the lowest LL \cite{Jain_book}. The gaps predicted from wave function energetics compare well with experiment as well \cite{GMP-exp1, GMP-exp2, GMP-exp3, Scarola2000}.  These collective modes have also been incorporated as central features of composite fermion effective field theories of lowest LL FQH states \cite{Lopez1993d,Murthy2003a}.  Unfortunately, accurate and well tested wave functions of the quasiparticle-quasihole excitations at 5/2 are still lacking.

A very useful approximation to FQH collective mode theory was originally formulated by Girvin, MacDonald and Platzman (GMP) ~\cite{GMP}.  It only requires a ground state ansatz and captures the fact that these collective modes are akin to density waves. The GMP excitation is gapped and neutral as well. In the long-wavelength limit, the GMP mode behaves as a spin-$2$ mode analogous to a graviton~\cite{Golkar2016, Yang2016, YangBo2012}. This graviton mode is chiral with its chirality (or polarization) depending on the FQH state~\cite{Liou-graviton}. 

The GMP approximation was used to study the low lying excitations of the $5/2$ state~\cite{Park2000, Wright2012}. Moreover, the GMP modes in Pfaffian, anti-Pfaffian, and PH-Pfaffian states have different polarizations~\cite{Haldane-Raman, Son-Raman}. As reviewed in Sec.~\ref{sec:future}, this feature is useful in probing the nature of the $5/2$ state. An additional type of collective excitation from Cooper-pair breaking was predicted for the Pfaffian state~\cite{Greiter1991}, which is dubbed as the neutral fermion mode~\cite{Bonderson-nf2011, Moller-nf2011}. It is gravitino-like and has spin $3/2$ in the long-wavelength limit~\cite{YangBo2012}. In the Pfaffian state, the GMP and neutral fermion modes can be viewed as superpartners~\cite{superspace}. It is expected that neutral fermion modes also exist in other non-Abelian candidates in the 16-fold way. However, no related study has been reported.

\section{Overview of numerical results} \label{sec:numerics}

Numerical support for the Laughlin state and composite fermion theory of the FQH effect in the lowest LL was definitive and set an extremely high standard. For example, finite size effects are reduced due to the large energy gaps and fast spatial decay of correlations.  Wave function overlaps between exact Coulomb ground states and the composite fermion states (including the Laughlin state) approach unity~\cite{Jain_book, Prange-book}. Numerical support for the Pfaffian description of the FQH effect at 5/2, while impressive as reviewed in this section, is less conclusive.

Early numerical support for the  Pfaffian description of the 5/2 state was provided in the seminal work by~\cite{x1-Morf}. Over the years, numerical work has continued to play an important role. Here we briefly discuss some technical details regarding the  numerical approaches used in studies of the FQH effect before discussing important landmarks and results.

\subsection{Numerical methods and geometries}
\label{sec_numerical_methods}

The non-interacting basis states for the Hamiltonian consisting of only the electron-electron interaction, Eq.~\eqref{eq:H}, are massively degenerate and there is no small parameter or normal state around which a perturbation theory can be formulated. An extremely useful parameterization, due to~\cite{Haldane_hierarchy} and discussed in more detail in \cite{Prange-book}, reformulated the Hamiltonian into so-called Haldane pseudopotentials. Eq.~\eqref{eq:H} can be rewritten as
\begin{eqnarray}
\label{eq:edham}
H &=& \sum_m V_{m,\mathrm{two-body}}^{(n)} \sum_{i<j}\hat{P}_{ij}(m)\;,
\end{eqnarray}
where $\hat{P}_{ij}(m)$ is a projection operator onto two-particle states with relative angular momentum $m$ and the $V_{m,\mathrm{two-body}}^{(n)}$ are the Haldane pseudopotentials in units of $e^2/\epsilon\ell_0$. $V_{m,\mathrm{two-body}}^{(n)}$ is the interaction energy between two electrons with relative angular momentum $m$ projected into the $n$-th electronic LL (the $n=1$ corresponds to the second LL, sometimes called the first excited LL). 

To study the FQH effect at $\nu=5/2$, the simplest model Hamiltonian is Eq.~\eqref{eq:edham} with the $V^{(1)}_m$ corresponding to the Haldane pseudopotentials for the Coulomb interaction between electrons confined to the second LL. The pure Coulomb interaction is an idealized situation because the magnetic field strength is taken to infinity (or equivalently the electron mass going to zero), i.e., the case of infinite spacing between LLs or zero LL mixing. 

It turns out that LL mixing~\cite{Bishara_LLM,Peterson_LLM,Simon_LLM,Sodemann_LLM} and the finite thickness~\cite{Peterson2008,Peterson2008a} of the quasi-two-dimensional electron system are important effects for second LL physics which must be taken into account.  Finite thickness is typically parameterized  by the width of the quantum well $w/\ell_0\sim 2-3$ for GaAs samples [see \cite{AndoRMP1982,Stern1984,Zhang1986}] and LL mixing is parameterized by the ratio $\kappa$ of the Coulomb interaction energy $E_\mathrm{Coul}$ to the cyclotron energy $\hbar\omega_c=eB/mc$ and depends inversely on the square root of the magnetic field strength:
\begin{eqnarray}
\kappa = \frac{E_\mathrm{Coul}}{\hbar\omega_c}\approx\frac{2.5}{\sqrt{B}}\;.
\end{eqnarray}
The second approximation is specific to GaAs heterostructures. Thus, when $\kappa\ll 1$, LL mixing can be ignored and the ``pure" Coulomb Hamiltonian projected into the second LL is a good model. However, as discussed in Sec.~\ref{sec:old-exp} the FQH effect at 5/2 has been observed over a wide range of magnetic field strengths: $1\;\mathrm{T} \lesssim B\lesssim 30 \;\mathrm{T}$, 
corresponding to $2.5 \gtrsim \kappa\gtrsim 0.5$, hence, $\kappa$ is never vanishingly small and is often sizable.

Besides the realistic (or idealized) Hamiltonian relevant for experimental systems mentioned above, there is a model Hamiltonian due to ~\cite{Greiter1992}, which produces the Pfaffian  as an exact zero-energy ground state. This (artificial) three-body Hamiltonian is written as
\begin{eqnarray}
H_3 = \sum_{i<j<k}S_{ijk}\left\{\nabla^2_j[\delta(z_i-z_j)\nabla^4_k\delta(z_i-z_k)]\right\}
\label{eq:H3}
\end{eqnarray}
where $S_{ijk}$ symmetrizes over all permutations of $ijk$. The form of $H_3$ is due to the fact that the Pfaffian  vanishes whenever three electrons coincide~\cite{Greiter1991}, see Eqs.~\eqref{eq:MR} and~\eqref{eq:Pf-def}. Note that $H_3$ breaks particle-hole symmetry explicitly as does the Pfaffian wave function. The particle-hole conjugate of $H_3$, called $\bar{H}_3$, produces the anti-Pfaffian as an exact ground state~\cite{Apf_Lee2007, Apf_Levin2007}. 

Three geometries are commonly utilized in numerical studies and pertinent details are discussed by~\cite{Jain_book}.  The experimental system is  quasi-two-dimensional  with edges created with confining potentials.  For numerical convenience and simplicity this system is mapped to a quasi-two-dimensional one with periodic boundary conditions (the torus geometry), a rotationally symmetric system (the disc geometry), and a compact geometry where the electrons on the two-dimensional plane are mapped to the surface of a sphere with the perpendicular magnetic field  produced by a magnetic monopole placed at the center (the spherical geometry). There is also the cylinder geometry (a torus with a cut) used primarily in DMRG calculations. Each geometry has pros and cons. 

The torus geometry is perhaps the most straightforward~\cite{Chakraborty_book,Haldane1985b,Yoshioka_book}. The system is a parallelogram with  periodic boundary conditions. The LL degeneracy per unit area is $B/\phi_0$ and the filling factor is precisely defined as $\nu=\rho/(B/\phi_0)$.  The energy eigenvalues are labeled by pseudo-momenta $(k_x,k_y)$. This geometry has genus one (it has a single hole) and allows a topological ground state degeneracy greater than one that can be leveraged to identify the topological order of an exact ground state. 

In the spherical geometry \cite{Wu1976,Wu1977} the electrons are placed on the surface of a sphere of constant radius $R=\sqrt{Q}\ell_0$ where $Q$ is the strength of the Dirac magnetic monopole at the center of the sphere. The total magnetic flux through the surface of the sphere is $2Q \phi_0$ and $2Q$ is an integer due to  Dirac's quantization condition.  The relationship between the number of electrons and the magnetic flux is given by
\begin{eqnarray}
\label{eq:2Q}
2Q=\nu^{-1}N - S
\end{eqnarray} 
where $S$ is an order-one number called the topological shift~\cite{Wen-Niu-shift,Wen-Zee-shift}.

Different shifts $S$ correspond to states with different topological orders. For example, the Pfaffian has $S=3$ while the anti-Pfaffian has $S=-1$, see Table~\ref{tab:summary}. The filling factor in this geometry is then defined as $\nu=\lim_{Q\rightarrow \infty} N/2|Q|$ and energy eigenstates are given as functions of total angular momentum $L$. This geometry has no edges and a finite degeneracy of $2(Q+n)+1$ for the $n$-th LL. This makes the sphere especially convenient to study bulk physics, ground states and low-energy excitation spectra, and ansatz wave functions. 

In the disc geometry, the single-particle states are especially simple~\cite{Jain_book}. In fact, the disc geometry is utilized often when representing ansatz wave functions in real space and when calculating Haldane pseudopotentials of various interaction potentials. The disc geometry requires a confining potential to keep the electrons from repelling one another off to infinity. Since the disc geometry necessarily has an edge it can be used to study edge physics.

While all the numerical work has finite size effects due to the modest number of electrons that can be studied, i.e., up to about $N=18$ for realistic Hamiltonians discussed in detail in Sec.~\ref{sec:numericsLLM} which has a total $L_z=0$ subspace dimension of over 29 million states, the energy gap necessary for a FQH state to exist reduces the finite size effects substantially. See the Supplemental Materials of~\cite{Balram2020} for the largest systems exactly diagonalized to date. We note that Monte Carlo simulations, which can treat on the order of hundreds of electrons, mostly involve the analysis of ansatz wave functions and not exact eigenstates.

\subsection{Spin polarization}
The FQH effect at 5/2 remained puzzling for a dozen years after its experimental discovery by ~\cite{Willett1987}. Initial ``tilt-field" experiments showed a disappearing FQH plateau upon tilting the sample with respect to the magnetic field direction while keeping the filling factor fixed. It  was initially thought that the ground state at 5/2 was spin-unpolarized because of this which largely ruled out non-Abelian physics and the Pfaffian state, see Sec.~\ref{sec:spinpolarization}. However, ~\cite{x1-Morf}'s seminal work changed the paradigm by providing substantial support for the Pfaffian description of 5/2. 

\cite{x1-Morf} studied the pure Coulomb interaction via exact diagonalization in the spherical geometry for up to $N=18$ electrons. As noted above, in the spherical geometry for electrons half-filling the second LL (the lowest spin-up and spin-down lowest LLs are taken to be fully occupied and inert) the relationship between $Q$ and $N$ for electrons in the second LL is $2Q=2N - S$. In the limit of $Q\rightarrow \infty$ the filling factor is $1/2$ in the second LL no matter the value of $S$. The  Pfaffian has $S=3$ but there is no a priori reason to fix $S$ in exact diagonalization\footnote{We note a technical point in numerical studies on the sphere. From Table~\ref{tab:summary} some of the shifts for various candidate states can be up to $S=9$. These large shifts are complicated to analyze due to an effect called ``aliasing" where an instance of $(2Q,N)$ in Eq.~\eqref{eq:2Q} can correspond to a different filling factor $\nu'$ and a different shift $S'$ or perhaps an excited state of an unrelated FQH state at yet another filing factor, see discussion in ~\cite{Morf2002}.}. However, any FQH candidate state has to be a  rotationally invariant state with $L=0$ (uniform density) and a finite energy gap.

\begin{figure}[htb]
\centering
\includegraphics[width=0.45\textwidth]{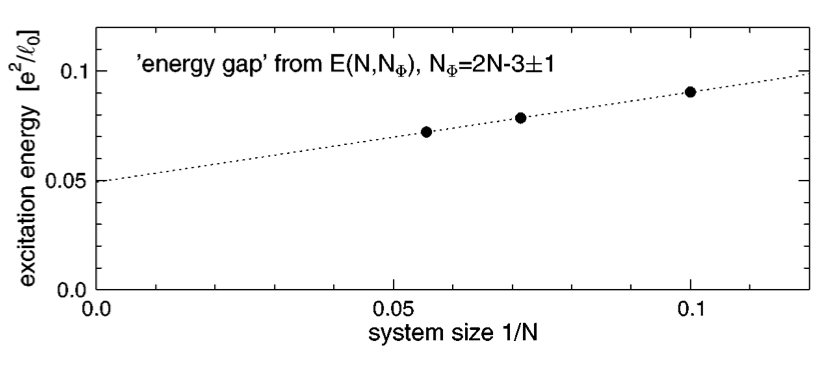}
\includegraphics[width=0.45\textwidth]{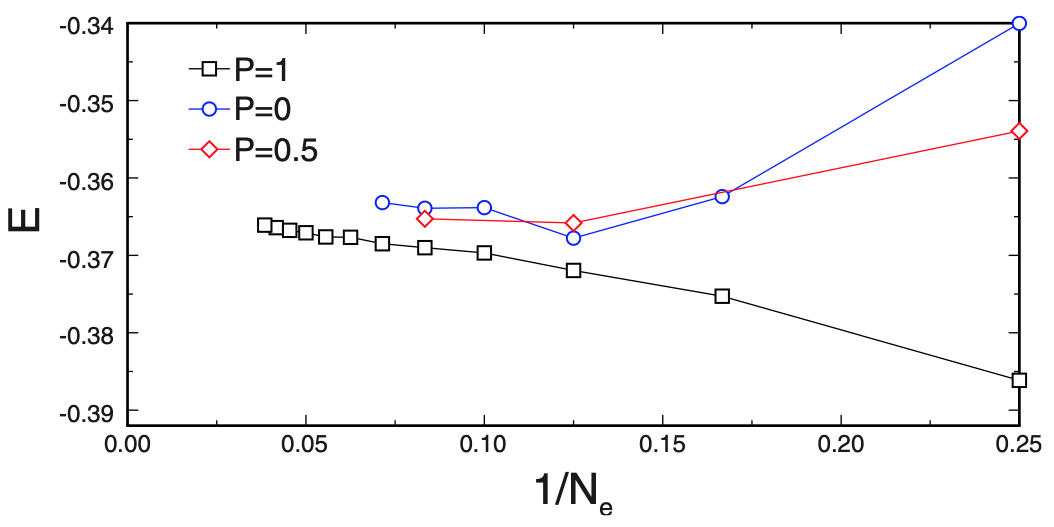}
\caption{(Top) \cite{x1-Morf} found the energy gap for the Pfaffian shift at $S=3$ to be finite in the thermodynamic limit and that the low energy dispersion for fixed number of particles $N$ displays a roton mode at finite wave vector (not shown). (Bottom) The ground state energy, obtained using density-matrix-renormalization-group techniques by \cite{num-spin-1} for Eq.~\eqref{eq:H} in the second LL in the spherical geometry, versus inverse particle number. The thermodynamic limit is obtained for $1/N\rightarrow 0$.  DMRG was able to study the largest systems to date and found the fully spin-polarized $P=1$ ground state at the Pfaffian shift of $S=3$ to be lower in energy than partially polarized states $P\neq 1$. This result significantly bolstered earlier results of ~\cite{x1-Morf}.  The figures are taken from \cite{x1-Morf} and \cite{num-spin-1} with permission.}
\label{fig:park_jain}
\end{figure}

\cite{x1-Morf} showed three important things: fully spin polarized states had lower ground state energy than spin-unpolarized states, $S=3$ had $L=0$ ground states for all even $N$ compared to other $S$, the state for $S=3$ was found to have a  finite energy gap in the thermodynamic limit [Fig.~\ref{fig:park_jain} (top)]. In addition, it was  shown that the wave function overlap between the exact ground state for $S=3$  and the Pfaffian state was substantial (60--80\%) and could be made nearly unity by increasing the $V_1^{(1)}$ pseudopotential by 10\% compared to the bare value --- the energy gap was also found to be maximum at this value of increased $V_1^{(1)}$.  

A technically important point for numerical calculations is that $H_3$, which in the spherical geometry is written as
\begin{eqnarray}
H_3 = \sum_{i<j<k}\hat{P}_{ijk}(3Q-3)\;,
\end{eqnarray}
where $\hat{P}_{ijk}$ is a projection operator onto electron triplets with maximum angular momentum $L=3Q-3$, can be solved to produce the Pfaffian state exactly in the second quantized representation. This greatly facilitates overlap calculations since the inner product can be calculated as a sum rather than a multi-dimensional integral. Furthermore, the energy spectrum of $H_3$ provides a model spectrum of a paired FQH state in a half-filled LL.

Note that ~\cite{x1-Morf} did not study the anti-Pfaffian with $S=-1$, which was not known at the time. Due to the particle-hole symmetry of the Hamiltonian, all the results would be identical\footnote{For a particle-hole symmetric Hamiltonian in the spherical geometry in a half-filled LL, all spectra for $S$ are identical to those with $S\rightarrow 2-S$. Hence, the Pfaffian spectra for $S=3$ is identical to the anti-Pfaffian spectra for $S=-1$.} to $S=3$ anyway. 

The spin-polarization of the Coulomb interaction in the second LL is a numerically taxing problem to study, i.e., the Fock space is formidably large. While \cite{x1-Morf}'s early calculations strongly supported a spin-polarized ground state for $S=3$, this result was more firmly established via DMRG for up to $N=26$ particle ~\cite{Feiguin-DMRG,num-spin-1}; see Fig.~\ref{fig:park_jain} (bottom). There is little numerical doubt that the FQH ground state at 5/2 is fully (or mostly) spin-polarized; see ~\cite{num-spin-2, Park1998} and \cite{num-spin-3} for more support using variational Monte Carlo techniques and Ginzburg-Landau theory, respectively. It is noteworthy that spin-polarization is maintained under realistic effects such as LL mixing and finite thickness of the quantum well~\cite{Rezayi2011}.

\subsection{Excitation gaps}

There are several energy gaps relevant for the FQH effect at the 5/2 state. The transport gap, the ``neutral" gap, and the neutral fermion gap. The thermodynamic stability of the FQH effect at 5/2 depends on all three.

\subsubsection{Transport gap}
The transport gap is the energy of a far separated quasiparticle/quasihole pair and is the gap experimentally measured by examining the longitudinal resistance; see Sec.~\ref{sec:old-exp}. Here we discuss the transport gap calculated via exact diagonalization in the spherical geometry. The transport gap can be calculated  two ways. One  is to calculate the ground state energy for  systems with $2Q=2N-3$ and $2Q=2N-3\pm 1$, the $\pm 1$ creates a two quasihole state or two quasiparticle state, respectively, with charges $\pm e/4$. The transport gap is the average of these excitation energies.  Another method is to consider the low-energy dispersion at fixed $N$ and take the gap between the lowest energy state for $L=N/2$ (for a paired state like the Moore-Read Pfaffian or $L=N$ for an unpaired state) and the $L=0$ ground state. 

Both of these methods for obtaining the transport gap are approximately the same. \cite{x1-Morf} found this transport gap to be approximately 0.05 $e^2/\epsilon\ell_0$ --- the $\nu=1/3$ Coulomb state in the lowest LL has a gap of about 0.1 $e^2/\epsilon\ell_0$ for comparison~\cite{Jain_book}. The gap calculations of ~\cite{x1-Morf} have been bolstered over the years, see~\cite{Feiguin-DMRG,Morf2002,new-num}.  We note that~\cite{Park2000} used the GMP approximation to calculate the energy gap dispersion for the Pfaffian state finding similar results to~\cite{x1-Morf} both qualitatively and quantitatively.

\subsubsection{Neutral gap}
The ``neutral" gap is defined as the difference between the $L=0$ ground state energy and the energy of the first excited state for any $L$.   ~\cite{x1-Morf} calculated the neutral gap at $2Q=2N-3$ and found it to be approximately 0.02 $e^2/\epsilon\ell_0$. This gap is thought to be accessible experimentally via photoluminescence spectroscopy; see Sec.~\ref{sec:raman}. In general, the neutral gap is smaller than, but of the same order of magnitude as, the transport gap. 

\subsubsection{Neutral fermion gap}

\begin{figure}[htb]
\centering
\includegraphics[width=0.45\textwidth]{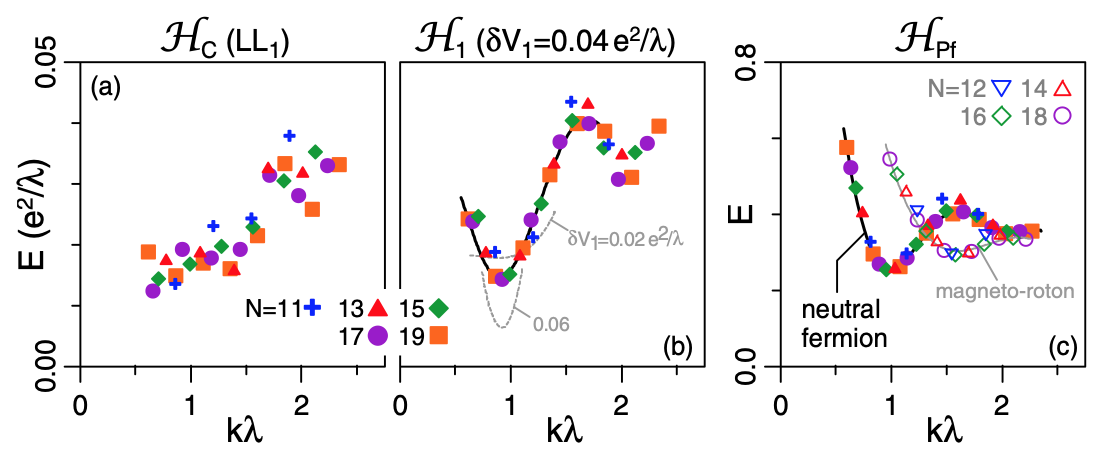}
\caption{Neutral fermion mode versus wave vector $k$ in units of the magnetic length calculated using exact diagonalization in the spherical geometry by \cite{Moller-nf2011}. The right panel shows the mode for the Pfaffian state (in this work they denote the second LL Coulomb Hamiltonian $H$ as $\mathcal{H}_\mathrm{C}(\mathrm{LL}_1)$ and $H_3$ as $\mathcal{H}_\mathrm{Pf}$), whereas the left panel is the neutral fermion mode for the $n=1$ Coulomb Hamiltonian. Increasing $V_1^{(1)}$ a little bit produces a mode closer qualitatively to that of the pure Pfaffian state, similar to overlaps and gaps found by \cite{x1-Morf}. Note that ~\cite{Moller-nf2011} denote $\lambda$ as the magnetic length $\ell_0$.  The figure is taken from \cite{Moller-nf2011} with permission.}
\label{fig:nfgap}
\end{figure}

The Pfaffian supports a neutral fermion excitation mode which is the gap for the Bogoliubov-de Gennes quasiparticles in the $p$-wave superconductor analogy for the Pfaffian state~\cite{Bonderson-nf2011,Greiter1991,Greiter1992,Moller-nf2011,MR1991,Read-Green} --- this is not the ``neutral" gap discussed above. This mode technically can be calculated in the spherical geometry by studying a system at  $S=3$ but for odd particle numbered systems instead of even  required for pairing. \cite{Bonderson-nf2011,Hutzel2019,Moller-nf2011,Sreejith2011b} calculated the neutral fermion mode for the Pfaffian state and for the ground state of the second LL Hamiltonian corresponding to 5/2, see Fig.~\ref{fig:nfgap}. Similar to the overlap and gap calculations of~\cite{x1-Morf}, increasing the $V_1^{(1)}$ Haldane pseudopotential by approximately 10\% produced a result more in line with the Pfaffian expectation. 

\subsubsection{Realistic effects on energy gaps}
Realistic effects such as LL mixing and/or finite thickness of the quantum well have been shown to generally reduce the size of the energy gaps by up to 50\%~\cite{Dean-gap-5/2,Luo2021,Morf2002,new-num,Peterson2008,Peterson2008a}. However, theoretical calculations still overestimate the experimentally measured energy gaps; see Sec.~\ref{sec:old-exp} below.

\subsection{Evidence for pairing}
\begin{figure}[htb]
\centering
\includegraphics[width=0.45\textwidth]{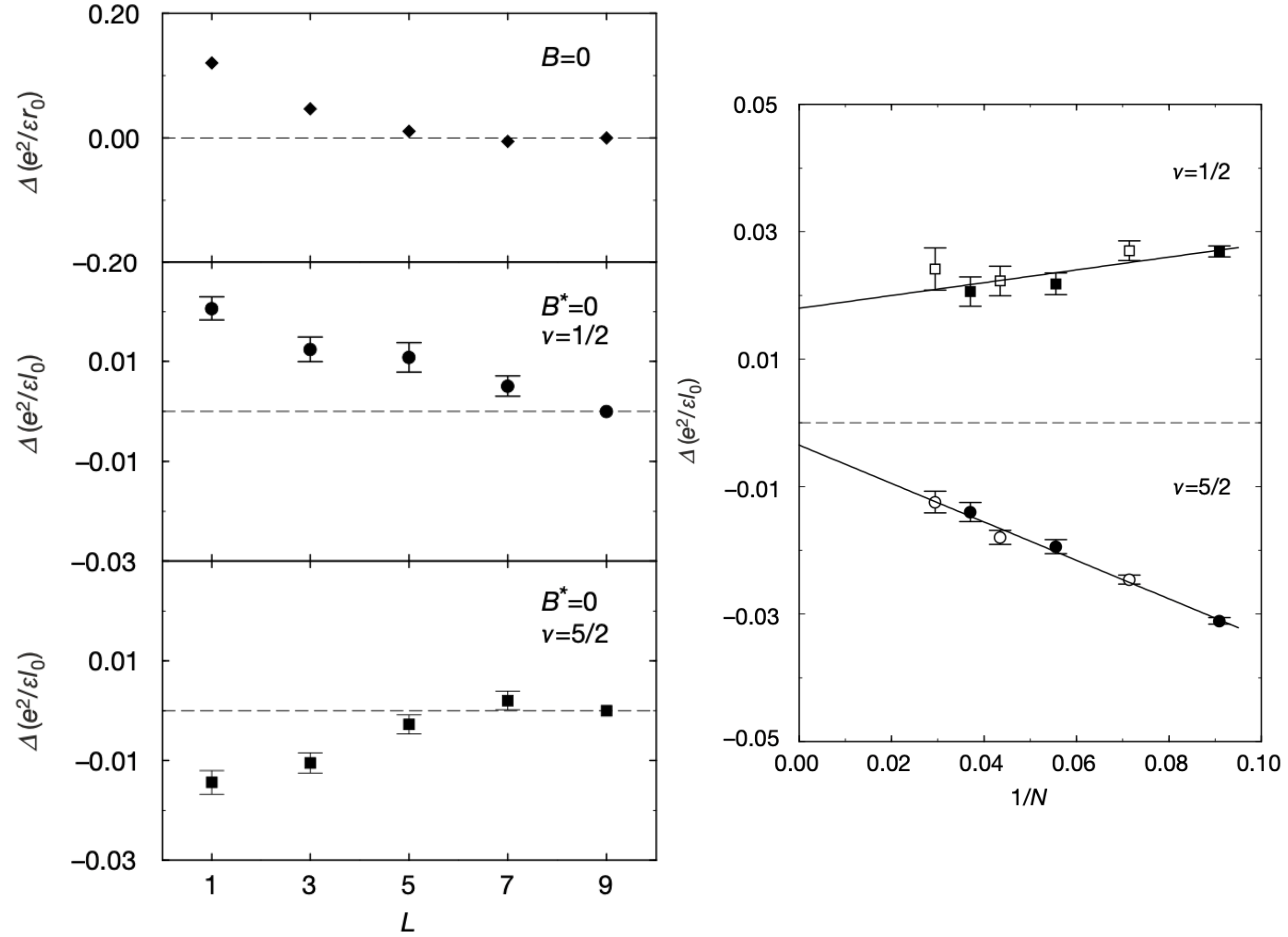}
\caption{(Left) The interaction energy for pairs of electrons ($B=0$) and composite fermions ($B^*=0$) in the lowest and second LL, respectively, versus angular momentum $L$, i.e., the pseudopotentials in the spherical geometry for $N=27$ particles. Increasing $L$ corresponds to increasing inter-particle separation when in a single LL. The short range composite fermion pseudopotentials are negative for 5/2 but not 1/2 where they are similar to those for electrons at $B=0$. (Right) The binding energy $\Delta$ for a composite fermion pair in the second Landau level, compared to the lowest Landau level, extrapolates to a negative value as $1/N\rightarrow 0$ indicating a pairing instability in the $L=1$ channel. Figures are taken from ~\cite{Scarola_nature} with permission.}
\label{fig:CFbinding}
\end{figure}

An important physical aspect of the Pfaffian state is that the composite fermions form pairs analogous to a $p$-wave superconductor. \cite{Scarola_nature} used Monte Carlo with the composite fermion wavefunctions in the spherical geometry, showing that the residual interaction between the composite fermions in the second LL is attractive. This attractive interaction is not found for composite fermions in the half-filled lowest LL, see Fig.~\ref{fig:CFbinding}.

\cite{LuSDSPark2010} showed an even-odd effect in the ground state energy per particle for the second LL Coulomb Hamiltonian indicative of pairing. This effect was absent in the lowest LL and instead was qualitatively similar to the ``shell" structure expected of a composite fermion Fermi sea. In addition, \cite{LuSDSPark2010} calculated a ``superconducting order parameter", $\langle c_*^\dagger(\mathbf{r}) c_*^\dagger(\mathbf{0})\rangle$, where the creation operators $c_*^\dagger$ are composite fermion creation operators since they are the objects that pair in the Pfaffian state compared to electrons, see Fig. 3 in \cite{LuSDSPark2010}.  In addition to the above calculations, a small increase in the pair correlation function of the Pfaffian state, calculated via Monte Carlo, for small distances has been interpreted as evidence for real space pairing~\cite{Park1998}.

The relative stability of the Pfaffian phase and further evidence for pairing was  demonstrated by ~\cite{Moller2008},  who considered a larger class of wave functions of the Pfaffian form, i.e.,
\begin{eqnarray}
\Psi_\mathrm{CF-pairing}({z_i}) = J^2\mathrm{Pf}\left[g(z_i-z_j)\right]
\end{eqnarray}
where $g(z_i-z_j)$ is an antisymmetric pairing function for the composite fermions. The Pfaffian state has $g(z_i-z_j)=1/(z_i-z_j)$, see Eq.~\eqref{eq:Pf-def}. By varying the form of the interaction potential and using the pairing function as a variational parameter, it was shown through the calculation of wave function overlaps in the spherical geometry between $\Psi_\mathrm{CF-pairing}$ and the exact ground state that the paring is maintained under realistic conditions, and the ground state remains in the Pfaffian phase. A similar conclusion was more recently obtained in the torus geometry by~\cite{Jain2021}.

\subsection{Topological properties}

Any candidate for the FQH state at 5/2 must have an energy gap.  There is also a consensus that some sort of pairing must be involved. However, the Pfaffian/anti-Pfaffian state has specific signatures related to yet another important property: non-Abelian topological order. 

\subsubsection{Topological shift}

The shift $S$ in the spherical geometry is a topological quantum number related to the topological order of the FQH state. The shift arises due to the orbital motion and the curvature of the sphere~\cite{Wen-Niu-shift}. Various candidate FQH states have an associated shift when represented in the spherical geometry. Incidentally the shift is related to the Hall viscosity~\cite{Read-viscosity}. As discussed above ~\cite{x1-Morf} found in the absence of LL mixing effects that $S=3$  was rotationally invariant for all even $N$ and had a finite energy gap. This is not necessarily true for other $S$.

The torus geometry does not have a shift.  However, the so-called infinite-DMRG calculations done on an infinite cylinder can cleverly extract the shift from the quantum entanglement data of the ground state as the leading term of a theoretical object called the ``momentum polarization"~\cite{Tu2013,Zaletel2013,Zaletel2015}.

\subsubsection{Ground state degeneracy}

In the torus geometry, the Pfaffian state displays an exact three-fold ground state degeneracy at unique values of pseudomomenta. Note that we have factored out the (trivial) two-fold center-of-mass degeneracy from the filling factor. This degeneracy can be understood analytically, see~\cite{Greiter1992}, and shows up numerically as well. Thus, if the second LL Coulomb interaction Hamiltonian is in the Pfaffian phase it should  show this three-fold (quasi)degeneracy. ~\cite{Peterson2008,Peterson2008a,Peterson-Papic-DS,Haldane-Pf2000,Haldane-Pf2009} all found the three-fold quasi-degeneracy manifest for the 5/2 system. In fact, ~\cite{Peterson2008,Peterson2008a} found that the realistic effect of finite thickness of the quantum well enhanced the degeneracy compared to the pure second LL Coulomb interaction.

\subsubsection{Wave function overlaps}
The numerical wave function overlap between the Pfaffian state and second LL Coulomb ground state was calculated by~\cite{x1-Morf} and found to be nearly 80\% and could be made nearly unity by increasing the short-range Haldane pseudopotential. We note that the interpretation of overlap calculations, while used widely in the study of FQH states, are subtle. Overlaps are necessarily calculated for finite size systems and vanish in the thermodynamic limit trivially. Thus, a high overlap should not be taken as evidence of the existence of a phase in isolation. Nonetheless, high overlaps between the exact ground state of second LL Coulomb interaction have been routinely calculated and have been shown to increase  with moderate finite thickness of the quantum well before eventually decreasing at very wide quantum wells~\cite{x1-Morf,new-num,Park2015,Papic-Regnault-DS,Peterson2008,Peterson2008a,Haldane-Pf2000,Rezayi2011,TylanTyler2015,Wang-PH-1/2}.

\subsubsection{Entanglement entropy and spectra}
The overlap is not universal and vanishes in the thermodynamic limit. The entanglement entropy and spectrum provide a more universal signature. A wave function's entanglement entropy is calculated by dividing the system into two subsystems $A$ and $B$, in orbital space\footnote{The division into subsystems can also be done in real space, see~\cite{Sterdyniak2012}. We consider only orbital cuts here.}, 
and constructing the reduced density matrix $\rho_A$ by taking a partial trace of all the degrees of freedom in the $B$ subsystem, or vice-versa~\cite{Srednicki1993}. The entanglement entropy is then the von Neumann entropy $S_E = - \mathrm{Tr}[\rho_{A}\mathrm{ln}\rho_{A}]$ of $\rho_A$. For gapped ground states, $S_E$ can be shown to generically scale with the length of the boundary between the two subsystems~\cite{Srednicki1993}. If a state has topological order, $S_E$ will show a reduction by a universal constant, i.e., for a topological state the entanglement entropy is expected to obey
\begin{eqnarray}
S_E = \alpha L - \gamma + \mathcal{O}(1/L)
\end{eqnarray}
where $L$ is the length of the boundary, $\alpha$ is a system dependent non-universal constant, and $\gamma$ is the so-called topological entanglement entropy~\cite{Kitaev2006,Levin-long_range}. For a topologically ordered state $\gamma=\ln\mathcal{D}$ where $\mathcal{D}$ is the total quantum dimension of the theory, see ~\cite{Das-Sarma}. For the Pfaffian, $\gamma=\ln\sqrt{8}$~\cite{Dong2008,Fendley2007,MR1991,Zozulya2007}.

Consistent with energy gaps and overlaps, \cite{Friedman2008} and \cite{Biddle2011} found that $\gamma$ for ground state of the half-filled second LL Hamiltonian was consistent with the value for the Pfaffian state when finite thickness of the quantum well was included, see Fig. 3 in that work.

\begin{figure}[htb]
\centering
\includegraphics[width=0.45\textwidth]{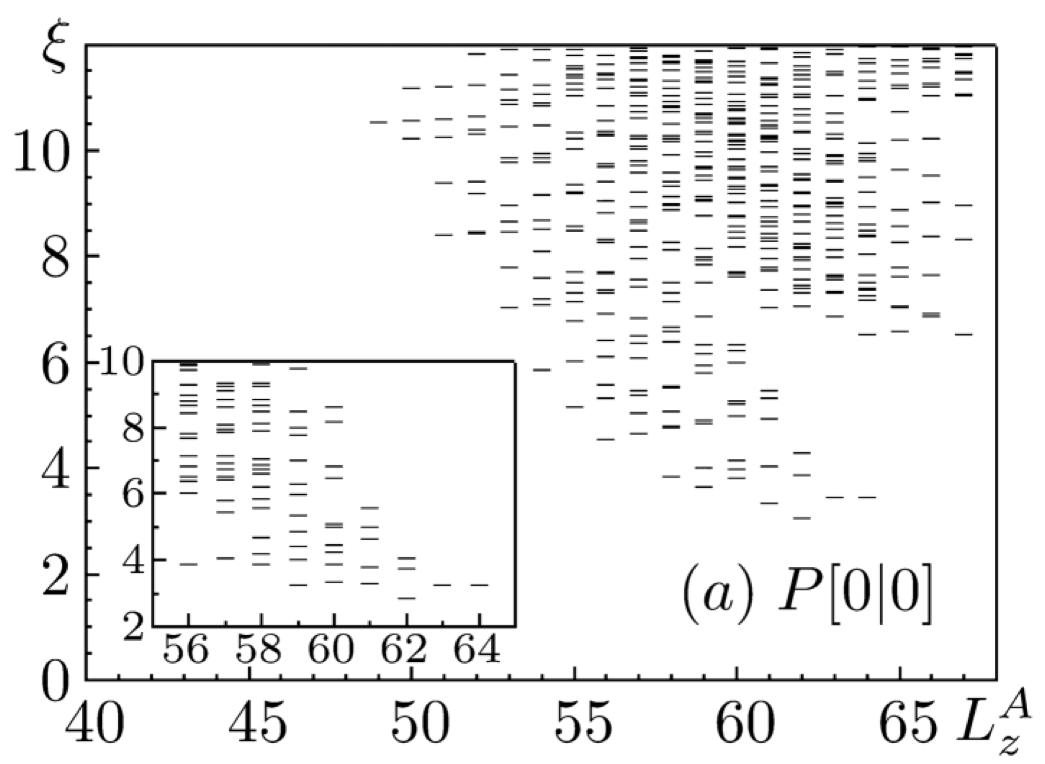}
\caption{The low-lying entanglement spectra $\xi$  for the ground state of the half-filled second LL Coulomb interaction Hamiltonian for $N=16$ electrons in the spherical geometry as a function of $z$-component of total angular momentum $L_z^A$ in the $A$ partition. The inset shows low-lying entanglement spectra for the Pfaffian state. The low-lying levels of both are in close correspondence and match the expectations from CFT for the Pfaffian. $P[0|0]$ corresponds to a particular orbital cut explained in detail by~\cite{Li2008}. The figure is taken from ~\cite{Li2008} with permission.}
\label{fig:ent_spect}
\end{figure}

It turns out there is far more information in $\rho_A$ than the entanglement entropy. ~\cite{Li2008} investigated the eigenvalues of $\rho_A$ and defined corresponding entanglement ``energies" $\xi_n$, whose spectrum is called entanglement spectrum. The entanglement spectrum can be used as a universal ``fingerprint" to identify the topological order of a wave function. The low-lying levels are related quantitatively and qualitatively to the related CFT  and its corresponding edge theory --- the orbital cut is thought to create an edge of sorts and the edge excitations can be studied. Fig.~\ref{fig:ent_spect} shows the entanglement spectra calculated by~\cite{Li2008} for the ground state of the half-filled second LL for a system with $N=16$ electrons in the spherical geometry. The inset shows the entanglement spectra for the Pfaffian wave function. For the Coulomb state there are several low-lying levels that match those of the Pfaffian and there is a so-called ``topological entanglement gap" to non-universal levels protecting the topological properties.

The low-lying entanglement spectrum for the 5/2 system is maintained when finite thickness effects are included in addition to LL mixing effects (Sec.~\ref{sec:numericsLLM}).

\subsubsection{Edge velocities}
While the entanglement spectrum is used to investigate the edge excitations by proxy, \cite{Wan-Pf2006} utilized the disc geometry to study the edge excitations of the Pfaffian state and the second LL Coulomb state directly. Chiral edge excitations were identified and fractionally charged excitations were consistent with the Pfaffian description. Additionally, the velocities for the fermionic and bosonic excitation branches $v_b$ and $v_f$, respectively, were found to satisfy $v_b \gg v_f$. In a follow up work, ~\cite{Yang-Pf2008} mixed in the three-body $H_3$ term perturbatively to the analysis (similar to LL mixing discussed below) and suggested that a smooth edge confinement potential might favor the anti-Pfaffian state while a sharp edge favors the Pfaffian. As mentioned above, finite size effects arise more readily in the disc geometry and this work required particular care to choose a proper confining potential. 

In contrast to edge studies in the disc geometry, ~\cite{Soule2013} studied edge modes of the Pfaffian state (compared to the second LL Coulomb ground state) in a QPC-type geometry using an open cylinder. With a combination of exact diagonalization, using $H_3$ to generate the Pfaffian state exactly, and Monte Carlo, it was found that $v_b\approx v_f$.

\subsubsection{Non-Abelian quasiparticles}

Perhaps the most interesting aspect of non-Abelian FQH states is the braiding statistics of the excitations and their non-Abelian nature~\cite{MR1991, Nayak1996, Prodan2009, RR1996}. Within the context of CFTs the non-Abelian braiding statistics are well-established as discussed above in Sec.~\ref{sec:theory}. In the spherical geometry the Pfaffian Hamiltonian in Eq.~\eqref{eq:H3} produces a zero-energy $2^{n-1}$ degenerate manifold of states from $2n$ quasiholes, separated from the continuum by an energy gap, when exactly diagonalized at $2Q=2N-3 + n$. The degeneracy $2^{n-1}$ is only obtained in the limit of $N\gg n$ and finite size effects produce more complicated degeneracies given in Table I of~\cite{RR1996} --- an example of the non-Abelian zero-energy quasihole manifold is shown Fig. 3 of the same work for a system with $N=10$ electrons. These quasihole excitations are analogous to the vortex excitations in the $p$-wave superconductor analogy of the Pfaffian state. Monte Carlo calculations were performed by ~\cite{Tserkovnyak2003} confirming the CFT predictions of the Pfaffian quasihole wave functions~\cite{Nayak1996}. Using the so-called bipartite composite fermion theory, \cite{Sreejith2011a, Sreejith2011b} constructed ansatz states for the non-abelian quasiparticle excitations that accurately represent the non-abelian excitations of Eq.~\eqref{eq:H3} and possibly are adiabatically connected to the low-energy states of the second LL Coulomb Hamiltonian.

However, a fair amount of numerical controversy continues regarding the question of whether non-Abelian excitations are manifest in a system with a realistic interaction appropriate for the FQH effect at 5/2. Generally, the Coulomb interaction, even when realistic effects such as LL mixing and finite thickness are included, does not show a zero-energy manifold of non-Abelian quasihole states separated from the continuum by an energy gap. Instead the quasihole degeneracy is strongly broken and the energy gap is not clearly discerned~\cite{PetersonPark2008,Toke2006,Wojs-Pf2010}. Recently, ~\cite{Hutzel2019} showed that the zero-energy quasihole manifold is adiabatically connected to a degeneracy-broken set of states of a two-body Hamiltonian related more closely to the second LL Coulomb Hamiltonian than the three-body Hamiltonian of Eq.~\eqref{eq:H3}. Thus, it is possible that the broken quasihole degeneracy generally observed in two-body interacting systems is a consequence of finite-size effects and mutual quasihole interactions~\cite{Das-Sarma}.

Finally, a recent parton wave function describing a half-filled FQH state, called the $\bar{2}\bar{2}111$ state was constructed based on the original idea in~\cite{Jain-221}. This wave function is constructed and analyzed in real space, via Monte Carlo for example, and was shown to be in the same phase as the anti-Pfaffian via an analysis of its entanglement spectra, wave function overlap, and pair correlation function~\cite{Balram-APf2018}. Since the anti-Pfaffian does not have a simple real space expression similar to Eq.~\eqref{eq:MR} for the Pfaffian, the $\bar{2}\bar{2}111$ state is a particularly convenient representation. Beyond the convenience of the $\bar{2}\bar{2}111$ representation of the anti-Pfaffian, the parton theory of the half-filled second LL suggests a unique hierarchy of expected FQH states. In particular, the sequence of states from $\nu=2+2/3$ to $2+1/2$ and $2+6/13$ discussed as possibly exotic in experiments \cite{Kumar2010}. The $2+6/13$ state, which is topologically equivalent to the hierarchy state of ~\cite{LevinHalperin2009}, is unexpected from weakly interacting composite fermion theory without the more robust  observations of states at $\nu=2+n/(2n+1)$ for $1\leq n\leq 5$, see~\cite{Balram2018b, Balram2019, Balram2021}. In fact, recent experimental work by ~\cite{PRX-graphene2022} in bilayer graphene used this method to investigate whether various half-filled FQH states arose from the anti-Pfaffian or the Pfaffian. 

\subsubsection{Adiabatic continuity}
Historically, an important piece of numerical evidence supporting the Laughlin description of the FQH effect at $\nu=1/3$ was that the Laughlin state was adiabatically connected to the Coulomb ground state. In this case the logic is straightforward. The Laughlin state for 1/3 is the exact ground state of Eq.~\eqref{eq:H} for which only $V_{1,\mathrm{two-body}}^{(0)}\neq 0$ while all others are set to zero, the so-called hard-core potential. To show adiabatic continuity between the lowest LL Coulomb ground state and the Laughlin state, one calculates the energy spectrum for the Coulomb Hamiltonian and then ``adiabatically" reduces all Haldane pseudopotentials other than $V_{1,\mathrm{two-body}}^{(0)}$ to zero. If the energy gap remains finite throughout this process then the ground states for the ``hard-core" potential and the lowest LL Coulomb Hamiltonian are equivalent.

For the FQH effect at 5/2, demonstrating adiabatic continuity is more complicated.  Complications arise because $H_3$ (producing the Pfaffian) is a three-body interaction that is (i) artificial and (ii) breaks particle-hole symmetry. Nonetheless, one can construct a Hamiltonian that interpolates between the second LL Coulomb interaction and $H_3$, i.e., $H(\alpha) = (1-\alpha)H + \alpha H_3$. In ~\cite{Storni-Pf2010} [and a related calculation in ~\cite{Hutzel2019}] it was shown that for the second LL Coulomb interaction the energy gap remained open as $\alpha$ was tuned from zero to one showing adiabatic continuity between the Pfaffian state and the ground state for 5/2, for finite sized systems. Interestingly, adiabatic continuity was not obtained for the lowest LL Coulomb interaction showing  second LL physics is crucial for the formation of the Pfaffian state.

\subsection{Competing phases and Landau level mixing}
\label{sec:numericsLLM}

So far we have discussed a preponderance of evidence showing that the Pfaffian (and anti-Pfaffian) states are viable states to describe the FQH effect at 5/2 even in realistic systems with finite quantum well thickness. However, small changes in Hamiltonian parameters have demonstrated that transitions to other competing states are possible~\cite{Peterson2008, Peterson2008a, Haldane-Pf2000, Wojs-Pf2010}.  For example, \cite{Haldane-Pf2000} performed exact diagonalization on both the sphere and torus and found a first-order quantum phase transition between the Pfaffian state and a compressible stripe (anisotropic) phase. Interestingly, the pure second LL Coulomb Hamiltonian was found to be barely on the compressible side of the phase transition but increasing the finite width of the quantum well was shown to drive the system into the Pfaffian phase.

Of particular interest are perturbations that can break the degeneracy between the Pfaffian and anti-Pfaffian. The system for the particle-hole symmetric second LL Coulomb Hamiltonian is  right at the cusp of a (first-order) transition between these degenerate states without a particle-hole symmetry breaking term~\cite{PetersonPark2008,Wang-PH-1/2}. Realistic effects such as finite thickness soften the interaction generally, i.e., the Haldane pseudopotentials are reduced. LL mixing, however, not only softens the interaction, it also explicitly breaks particle-hole symmetry~\cite{Bishara_LLM}. Since the Pfaffian and anti-Pfaffian are related via particle-hole conjugation, LL mixing is an important effect.

LL mixing can be taken into account in essentially two ways. One way is to expand the Fock space to include higher and lower LLs, i.e., allow basis states with partially filled lower and higher LLs instead of projecting all the physics to the half-filled second LL. This increased Fock space must still be truncated to a finite size in order to do any calculations. \cite{Papic2012,Rezayi2011,Zaletel2015} all analyzed the 5/2 FQH state using various non-perturbative methods that are not controlled or exact in any limit. 

In contrast, \cite{Peterson_LLM,Simon_LLM,Sodemann_LLM,Wooten2013} constructed an effective Hamiltonian that included LL mixing perturbatively to lowest order in the LL mixing parameter $\kappa$. Hence, this approximation is exact for  $\kappa\rightarrow 0$. Further, ~\cite{Peterson_LLM} constructed an effective Hamiltonian that included LL mixing and finite thickness. This latter effect is subtle since it requires LL and quantum well sub-band mixing. The form of the effective Hamiltonian is
\begin{equation}
\label{eq:Heff}
\begin{split}
H(w,\kappa) &= \sum_m \left[V_{m,\mathrm{two-body}}^{(1)}(w)\right.  \\
& +\left. \kappa\;\delta V_{m,\mathrm{two-body}}^{(1)}(w)\right]
\sum_{i<j}\hat{P}_{ij}(m)\\
&+\kappa \sum_m \tilde{V}^{(1)}_{m,\mathrm{three-body}}(w)\sum_{i<j<k} \hat{P}_{ijk}(m)
\end{split}
\end{equation}
where $V_{m,\mathrm{two-body}}^{(1)}(w)$ is the bare finite thickness dependent two-body Haldane pseudopotential, 
$\delta V^{(1)}_{m,\mathrm{two-body}}(w)$ is its perturbative correction due to LL and sub-band mixing, and $\tilde{V}_{m,\mathrm{three-body}}^{(1)}(w)$ is the emergent particle-hole symmetry breaking three-body pseudopotential~\cite{Simon2007}.

\subsubsection{Perturbative LL mixing treatment I}

\cite{Wojs-Pf2010} were the first to numerically investigate the particle-hole symmetry breaking effective LL mixing Hamiltonian constructed by~\cite{Bishara_LLM}. Assuming spin-polarized electrons, \cite{Wojs-Pf2010} studied the ground state and low-energy excitations of the realistic second LL Hamiltonian as a function of $\kappa$ at $S=3$ (for the Pfaffian) and $S=-1$ (anti-Pfaffian). It was found, primarily through wave function overlaps, that the Pfaffian and its low-energy excitations were favored compared to the anti-Pfaffian. We note, however, that the two-body pseudopotentials of the effective LL Hamiltonian calculated by~\cite{Bishara_LLM} were incorrect due to a subtle normal-ordering error in the perturbation theory --- this error was corrected by~\cite{Peterson_LLM,Sodemann_LLM}.

\subsubsection{Non-perturbative LL mixing treatment I}
\cite{Rezayi2011} did an exact diagonalization study in the torus geometry using a truncated Fock space that included LL mixing non-perturbatively. In particular, it was shown by comparing the size of the wave function overlap between the exact ground state and the Pfaffian or anti-Pfaffian that a model that included three LLs in the trucated Fock space, the Pfaffian was preferred.  However, increasing the Fock space to include four and five LLs, the anti-Pfaffian was preferred. 

We note that~\cite{Wojs2006} studied the 5/2 FQH effect under LL mixing using a truncated Fock space similar to~\cite{Rezayi2011} finding chiefly that the energy gap is strongly reduced. However, they did not discuss the anti-Pfaffian since it had not been proposed yet.

\subsubsection{Perturbative LL mixing treatment II}

\begin{figure}[htb]
\centering
\includegraphics[width=0.45\textwidth]{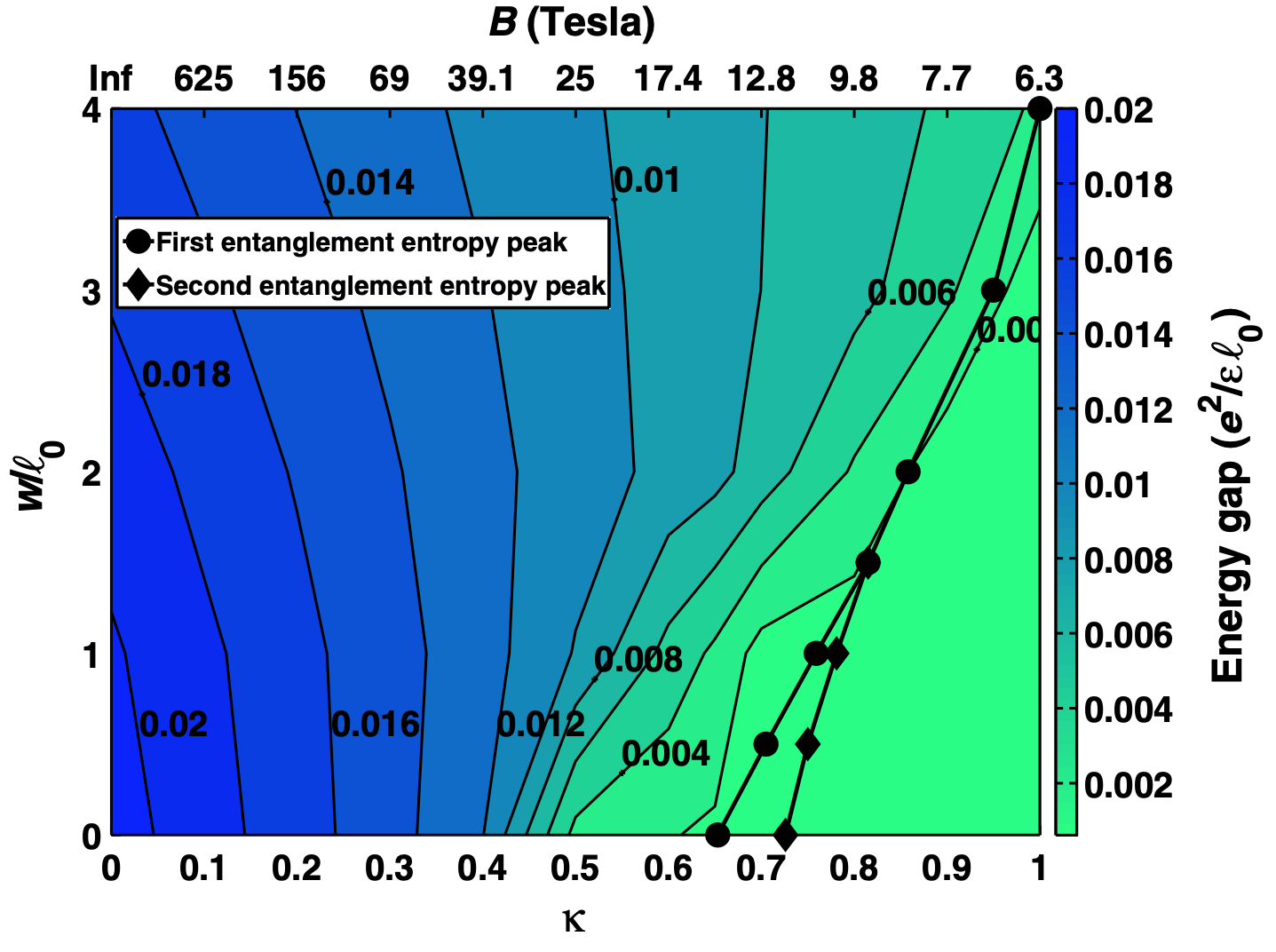}
\caption{Approximate quantum phase diagram for the FQH effect at 5/2 calculated by ~\cite{new-num} as a function of quantum well width $w/\ell_0$ and LL mixing parameter $\kappa$ in the spherical geometry for a system of $N=18$ electrons. The area left of the lines with black circles and diamonds corresponds to the Pfaffian state as determined by wave function overlap, entanglement spectrum, and a particle-hole symmetry breaking order parameter. The color indicates the strength of the ``neutral" energy gap which indicates the stability of the phase. The magnetic field values in Tesla are  shown on the upper ordinate ($\kappa\propto 1/\sqrt{B}$). The lines with the black circles and diamonds indicate peaks in the entanglement entropy marking phase transitions. The area to the right of the black diamonds is a non-FQH state. The small area in-between the circles and diamonds around $0.6<\kappa<0.7$ has a possible FQH ground state with edge modes moving opposite to the Pfaffian.  The figure is taken from ~\cite{new-num} with permission.}
\label{fig:52_qpd}
\end{figure}

The effective Hamiltonian in Eq.~\eqref{eq:Heff} was analyzed via exact diagonalization in the spherical and toroidal geometry by~\cite{new-num}. It is noteworthy that only the lowest $m\leq 8$ emergent three-body pseudopotentials $\tilde{V}_{m,\mathrm{three-body}}^{(1)}$ were taken into account by~\cite{new-num} --- above $m=8$ some of the three-body terms become multi-valued and it is non-trivial to consider these types of terms in calculations. However, the $\tilde{V}_{m,\mathrm{three-body}}^{(1)}$ decrease in value for increasing $m$ and presumably have little effect, see Sec.~\ref{sec:rezayiLLM} below to revisit this aspect. By analyzing wave function overlaps, entanglement spectra, energy gaps, and an order parameter used to determine particle-hole symmetry breaking, it was found that the Pfaffian state edges out the anti-Pfaffian throughout parameter space. The approximate quantum phase diagram calculated by~\cite{new-num} is shown in Fig.~\ref{fig:52_qpd}. The stability of the Pfaffian phase was measured via the ``neutral" energy gap. Sharp peaks in the entanglement entropy indicate phase transitions to a compressible state for $\kappa>0.7-1$ (depending on $w/\ell_0$).  Between about $0.6<\kappa<0.7$ there was a  hint of an intermediate FQH state. The entanglement spectra in this small window had low-lying levels with a slope opposite to the Pfaffian indicative of a state with edge modes opposite to the Pfaffian. However the entanglement and energy gaps were too small to reach a firm conclusion. Recent diagonalization and overlap calculations for $\kappa >1$ suggest a re-entrant anomalous quantized Hall state captured by Bose-Einstein condensates of composite bosons~\cite{Das2022}.

\subsubsection{Non-perturbative LL mixing treatment II}
\begin{figure}[htb]
\centering
\includegraphics[width=0.45\textwidth]{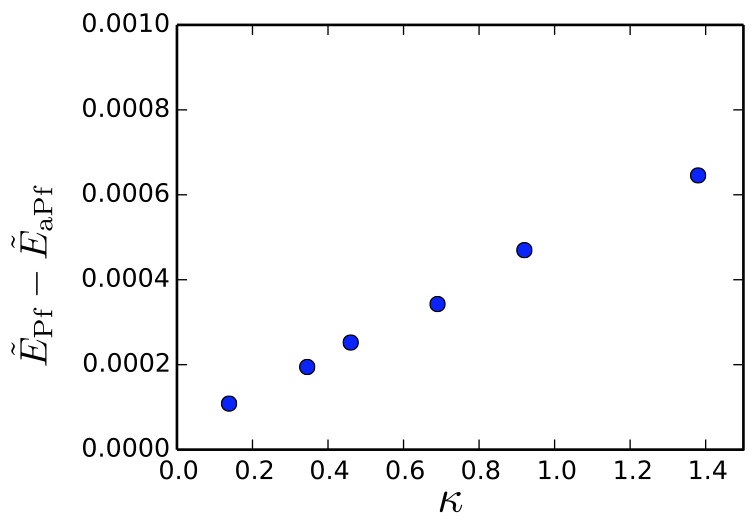}
\caption{Energy estimate from ~\cite{Zaletel2015} of the splitting between the Pfaffian and anti-Pfaffian states calculated via DMRG using a truncated Fock space simulating LL mixing as a function of $\kappa$. The positive value indicates the anti-Pfaffian is preferred from the experimentally relevant value of $\kappa$ towards zero.  The figure is taken from ~\cite{Zaletel2015} with permission.}
\label{fig:dmrg_aPf}
\end{figure}

In contrast to the perturbative approach of ~\cite{new-num}, ~\cite{Zaletel2015} utilized infinite-DMRG to explore LL mixing effects within a truncated Fock space including up to five LLs. Starting at an experimentally relevant value of $\kappa=1.38$, infinite-DMRG was performed on a truncated system, at zero quantum well thickness, with three LLs for decreasing values of $\kappa$ toward the perturbative limit of $\kappa=0$. The infinite-DMRG was performed by incorporating the matrix product form of either the Pfaffian or anti-Pfaffian on an infinite cylinder with a finite radius of approximately 20$\ell_0$. By analyzing the energy splitting between the Pfaffian and anti-Pfaffian (see Fig.~\ref{fig:dmrg_aPf}), they found the the anti-Pfaffian phase dominates for all $\kappa$. 

\subsubsection{Perturbative LL mixing treamtment III}
\label{sec:rezayiLLM}

Motivated by the contrasting results of ~\cite{new-num} and ~\cite{Zaletel2015}, ~\cite{Rezayi-PRL} re-investigated the perturbative approach for zero quantum well width using the torus and spherical geometries for $N=15$ and 16 electron systems. The torus geometry was utilized to directly compare the Pfaffian and anti-Pfaffian since there is no shift. The main difference between ~\cite{Rezayi-PRL} and ~\cite{new-num} was the inclusion of an approximate~\footnote{The multi-component $m=9$ three-body term was diagonalized and the dominant term was taken as a single $m=9$ term in Eq.~\eqref{eq:Heff}.}  $m=9$ three-body LL mixing pseudopotential term in Eq.~\eqref{eq:Heff}.  As noted above, most of the emergent three-body pseudopotentials due to LL mixing above $m=8$ are multi-component. Including this term tipped the balance toward the anti-Pfaffian as determined by wave function overlaps between the exact ground state and the Pfaffian or anti-Pfaffian, in both geometries, up to $\kappa\sim 1$ where the overlap with the anti-Pfaffian vanishes (the Pfaffian overlap collapsed early around $\kappa\sim 0.6$). The ``neutral" energy gap was also shown to be larger for the anti-Pfaffian compared to the Pfaffian state. This result was found to be rather stable to changes in the value of the $m=9$ three-body pseudopotential. However it is unknown to what extent quantum well sub-band mixing might affect the result.

\subsubsection{Perturbative LL mixing treatment IV}

In~\cite{TylanTyler2015} a phase diagram was calculated via a perturbative LL mixing treatment and edge effects were analyzed in the disc geometry. Here both LL mixing and the neutralizing background break the Pfaffian/anti-Pfaffian degeneracy. It was found there was a quantum phase transition from anti-Pfaffian to Pfaffian as $\kappa$ is increased from zero, qualitatively in contrast to what was found by~\cite{new-num}. In the thermodynamic limit, the boundary effect (from the trapping potential) that breaks the Pfaffian/anti-Pfaffian degeneracy is negligible as compared to the LL mixing effect from the bulk. Meanwhile, the actual sample does have a finite size where the edge effect is present, which may play a role in breaking this degeneracy.

\subsection{Unexpected experimental results and the PH-Pfaffian}

Numerical work has demonstrated that the Pfaffian/anti-Pfaffian is a strong candidate for a realistic model for the FQH effect at 5/2 that includes finite thickness of the quantum well. LL mixing effects allow the possibility to determine whether a realistic system prefers the Pfaffian or anti-Pfaffian by explicitly breaking particle-hole symmetry. As discussed above, numerically that question remains unanswered as some calculations suggest the Pfaffian while others suggest the anti-Pfaffian. To some extent this question will have to be settled experimentally.

As discussed below in Sec.~\ref{sec:thermal-Hall}, thermal conductance experiments by~\cite{Banerjee2018} were recently performed and did not settle the Pfaffian/anti-Pfaffian question. Instead, these experiments were consistent with a candidate state that is particle-hole symmetric known as the PH-Pfaffian~\cite{Zucker2016}; see Eq.~\eqref{eq:PHPf} above. In the spherical geometry, this state has a shift of $S=1$ (the particle-hole symmetric point) and has scant numerical support at this time. The second LL Coulomb interaction does not consistently obtain a rotationally invariant state with $L=0$ for all $N$ at $2Q=2N-1$ with or without realistic effects of finite width and LL mixing. Furthermore, the overlap between the PH-Pfaffian and the exact second LL Coulomb ground state was shown by ~\cite{Balram-APf2018} to be nearly zero. On the other hand, the overlap with the lowest-lying $L=0$ LLL Coulomb state is rather high leading ~\cite{Balram-APf2018} to speculate that the PH-Pfaffian is a critical state corresponding to a particle-hole symmetric pairing instability of the composite fermion Fermi sea [see also the theoretical discussion in~\cite{Milovanoic-pair}]. These results are consistent with those of~\cite{Geraedts2016,Mishmash,Pakrouski2021,Rezayi2021,Yutushi2020}. 

There are theoretical suggestions by ~\cite{Milovanovic_LLM, Milovanovic2021, Milovanovic2020, Zucker2016} that LL mixing and/or disorder are needed to stabilize the PH-Pfaffian state. Recently, \cite{Zhu-Sheng} studied disorder-driven transitions in the 5/2 FQH effect and found a transition from a pure Pfaffian/anti-Pfaffian phase, to an (unknown) intermediate phase, to a compressible composite fermion Fermi sea phase. The intermediate phase is consistent with either the PH-Pfaffian or the Pfaffian/anti-Pfaffian puddle state~\cite{Lian2018,Mross2018,Wang2018,Zhu-domain}. Finally, \cite{Luo2017} found a positive gap (albeit less than the one for the Pfaffian shift of $S=3$) for the PH-Pfaffian shift of $S=1$ at finite $\kappa$ via exact diagonalization on the sphere for $N=10$ electrons.  In this work they utilized yet another method of approximating LL mixing using an effective renormalized electron-electron interaction taking into account the polarizability of all the other Landau levels (random phase approximation) which preserves particle-hole symmetry.

\subsection{Summary of numerical results}

Extensive numerical work has been done since the discovery of the FQH effect at $\nu=5/2$. The preponderance of evidence support the Pfaffian/anti-Pfaffian description for a realistic second LL Hamiltonian that takes into account the finite width of the quantum well. There is numerical support for the notion that the half-filled second LL Coulomb system is close to a first-order phase transition that breaks particle-hole symmetry between the Pfaffian and anti-Pfaffian. LL mixing is a realistic effect that explicitly breaks particle-hole symmetry. This effect has been incorporated into numerical studies in both the perturbative limit (exact as $\kappa\rightarrow 0$) and non-perturbative limit with mixed results. At the time of this writing, the (slim) majority of studies conclude the anti-Pfaffian is favored by LL mixing effects. 

Meanwhile, thermal Hall conductance measurements at $\nu=5/2$ yielded a result that was consistent with neither the Pfaffian nor anti-Pfaffian. Instead the result points toward either the PH-Pfaffian or disorder-induced Pfaffian/anti-Pfaffian puddle formation. The PH-Pfaffian has scant numerical support and disorder is notoriously difficult to include in numerical studies.

\section{Early experiments and evidence for/against non-Abelian states} \label{sec:old-exp}

Theoretical developments in the quantum Hall effects walk hand-in-hand with experimental discovery and verification of hypotheses. Given the long list of possible candidates to describe the $5/2$ state (Table~\ref{tab:summary}),  it is natural to rely on experiments to distinguish these candidate theories.  Of central interest is whether or not experiments can verify the non-Abelian nature of the 5/2 state, and if so, which particular type.  In this section, we summarize early experimental methods used to probe various properties of the 5/2 state.  We review these results in the context of verification or exclusion of assumptions underlying certain candidate states. The next section focuses on reviewing more recent results from thermal Hall conductance measurements specifically. See also prior reviews of experimental progress in~\cite{Sarma1997, Heiblum-Feldman, 5/2-review-2019, Lin-review, Schreiber2020, Willett-review}. 

\subsection{Transport and optics: bulk energy gap}

Incompressibility in quantum Hall fluids requires a bulk energy gap.  Larger gaps define more robust states and a vanishing of the gap defines phase boundaries in parameter space.  Gaps at different momenta have acquired specific terminology. 

The``transport gap" refers to the energy to create quasiparticle/quasihole pairs that are separated much further than their size, usually assumed to be directly related to gap inferred from the thermal activation transport measurements (in the following we refer to the activation gap measured in experiment as the``transport gap" to make connection with expectations from theory discussed in Sec.~\ref{sec_numerical_methods}).  The ``neutral" bulk gap refers to the energy needed to create a quasiparticle-hole pair that are separated on the order of their size.  

Numerical work shows that the quasiparticles at $\nu=5/2$ can be as large as $\sim 10 \ell_0$ \cite{Morf2002} and that the transport gap is expected to be much weaker than it is for filling factor 1/3 (see Sec.~\ref{sec:numerics}). The small gap obtained from theory is consistent with measurements showing precise quantization of the 5/2 plateau in the Hall resistance only at much lower temperatures than other fractions \cite{Pan1999b}.  The parameter dependence and size of these gaps offer a direct route to compare theory and experiment.  Engineering a larger gap is an important goal for observing non-Abelian quasiparticles because many experiments become technically easier with large gaps. 

Detailed transport measurements have been performed on the 5/2 state since its discovery.  The transport gap was typically found to be less than $\approx600$~mK although a recent breakthrough set a record at $\approx820$~mK \cite{Chung2021}. As discussed below, the gap depends on sample parameters.  The measured transport gap of the 5/2 state is consistently found to be well below theory~\cite{Choi2008, Dean2008, Nuebler2010, Samkharadze2011} [as with the other FQH effect states \cite{Sarma1997, Jain_book}].  

Several factors are known to lower the measured transport gap in comparison to the theoretical estimate of the charged gap.  Finite thickness of the two-dimensional electron gas and LL mixing both soften the short range part of the Coulomb interaction and therefore lower the theoretical estimates \cite{Sarma1997, Jain_book}.  But discrepancies still remain that are typically ascribed to disorder, which are seldom included in theoretical estimates of the gap \cite{Morf2003, Yang-mobility}.  

Experimental parameters have shown a wide tunability of the transport gap.  The following reviews experimental results showing the behaviour of the transport gap at 5/2 as various parameters are tuned.  These include density, in-plane fields, quantum well width, disorder, and other aspects of sample design.

\subsubsection{Density dependence}

Electronic gating of the two-dimensional electron gas allows tuning of its density, $\rho$.  At fixed filling, increasing the density requires a linear increase in the magnetic field since $\rho=\nu B/\phi_0$.  Since the Coulomb energy scales as $1/\ell_0\propto\sqrt{B}$, increasing the density would therefore nominally increase the strength of the 5/2 energy gap in comparison to other energy scales, e.g., the temperature.   

Early studies at 5/2 showed a smooth dependence of the transport gap on density~\cite{Pan2001}.  But the role of density has been more widely explored~\cite{Liu2011, Pan2014, Reichl2014, Samkharadze} to reveal more complex behavior. It has been found that the transport gap of the 5/2 state is maximized at certain densities~\cite{Liu2011, Reichl2014} due to competing effects at high densities that include sub-band mixing.  Furthermore, \cite{Samkharadze} found anomalous behaviour in the the transport gap and a possible topological phase transition at low densities, where LL mixing may play a role.  At even lower densities, the gap was found to vanish in an apparent spin transition \cite{Pan2014} with later evidence for a nematic phase~\cite{Samkharadze2016, Schreiber2020, Schreiber2018} (see Sec.~\ref{sec_tilt}). Competing phases therefore also play an important role in the density dependence of transport gap measurements at 5/2. 

Fig.~\ref{fig:mobilirtyvsdensity} plots some of the measured transport gaps at 5/2 reported in the literature (before 2014) as a function of the density of the two-dimensional electron gas.  The data do not show a clear increasing trend with density.  While competing phases play a role, other important effects such as impurity screening also significantly impact gap measurements. 

\subsubsection{Disorder, mobility, and sample design} 

Disorder is believed to be responsible for a lowering of the transport gap.  Disorder is typically parameterized by a constant shift $\Gamma$, where the measured transport gap is compared with $\Delta_{\text{A}}-\Gamma$ and $\Delta_{\text{A}}$ is the transport gap in the absence of disorder. The microscopic origin of $\Gamma$ is not completely understood.  Example scattering mechanisms include background impurities, interface roughness, alloys, or even phonons \cite{Ahn2022}.  Generally speaking, disorder which is in some way biased to lower the energy of spatially non-uniform excitations in comparison to otherwise uniform FQH ground states, would tend to lower the energy gap.   

Mobility is often used as a proxy for sample quality because increases in zero-field sample mobility tend to, on average, increase the measured transport gap at 5/2  \cite{Choi2008, Chung2021, Dean-gap-5/2, Deng2014, Pan2011, Qian2017b, Reichl2014, Samkharadze, Samkharadze2011, Shingla2018}. But it is well known that the interplay of disorder, length scales, screening, and temperature leaves the zero-field sample mobility as only rather indirectly related to sample quality at it pertains to 5/2~\cite{Ahn2022, DasSarma2014, DasSarma2014a}.  This is consistent with a recent comparison between low field electron quantum lifetime measurements and 5/2 transport gaps which found a lack of correlation between these two quantities \cite{Qian2017}.  The energy gap of the 5/2 state, however, was found to depend monotonically on the mobility in the regime in which only one scattering mechanism dominates, in this case alloy scattering~\cite{Deng2014, Kleinbaum2020}. 

\begin{figure}[htb]
\centering
\includegraphics[width=0.45\textwidth]{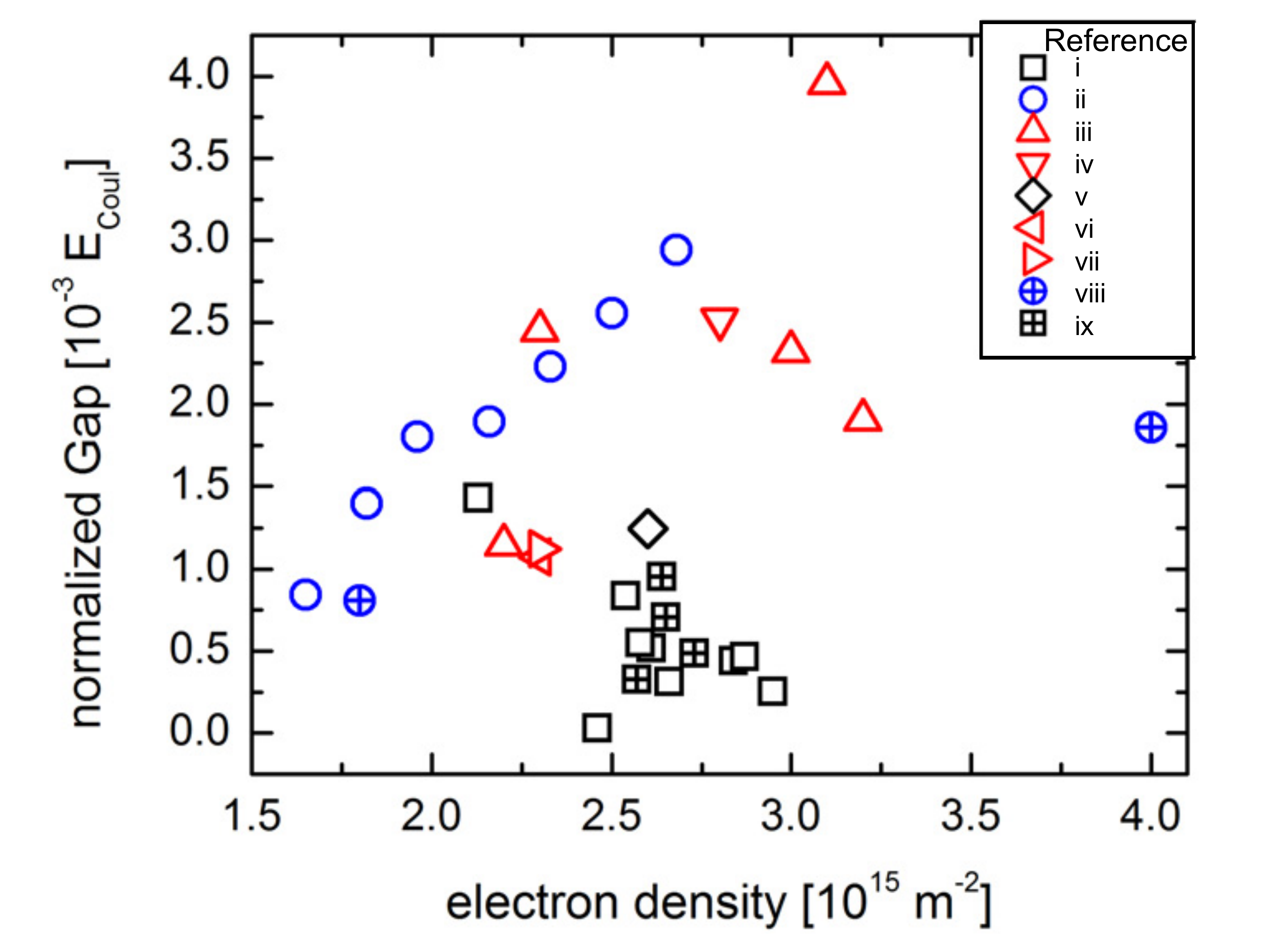}
\caption{Transport gaps measured at filling 5/2  normalized to Coulomb energy units ($\text{E}_{\text{Coul}} =e^2/\epsilon \ell_0)$ as a function of density of the two-dimensional electron gas as reported for various  heterostructures in the literature.  The symbols correspond to the following references: i) \cite{Reichl2014}, ii) \cite{Nuebler2010}, iii) \cite{FQH19/8_2}, iv) \cite{Choi2008}, v) \cite{Miller2007}, vi) \cite{Eisenstein1990}, vii) \cite{Pan1999}, viii) \cite{Pan2011}, and ix) \cite{Gamez2013}. The figure is taken from \cite{Reichl2014} with permission.}
\label{fig:mobilirtyvsdensity}
\end{figure}

Screening of sample dopants and impurities can help increase the 5/2 transport gap.  Fig.~\ref{fig:mobilirtyvsdensity} shows a general increase in the gap with density (as nominally expected) but there is not a clear trend because many other factors impact the gap.  For example, moving dopants far from the two-dimensional electron gas was reported to increase the 5/2 gap \cite{Reichl2014}.   Also, a non-trivial dependence was found depending upon the range of disorder in the type of sample used \cite{Pan2011}.  Furthermore, the FQH effect was observed at 5/2 at rather low mobility after illumination \cite{Gamez2013}.  These findings underscore the complex role of impurity screening on sample quality.  

The non-trivial interplay of impurities and other parameters such as sub-band splitting led to work on the role of sample design itself to possibly engineer higher 5/2 transport gaps. It was found that samples designed to lower impurity concentrations while allowing a wide tuning of the electron density maximizes the 5/2 gap~\cite{Chung2021, Liu2011, Pan2011, Reichl2014, Watson2015}.  Furthermore, while it is usually expected that widening the width of quantum wells lowers the transport gap (due to a softening of the short range part of the effective Coulomb interaction; see Sec.~\ref{sec:numerics}), some experiments do show that widening the well width can increase the gap~\cite{Xia2010}, consistent with numerical simulations incorporating the trade off in competing phases~\cite{Peterson-review}.  

A recent breakthrough increased the 5/2 gap in GaAs-based quantum wells to a record 820mK~\cite{Chung2021}.  This significant ($30\%$) improvement over the previous record~\cite{Choi2008, Qian2017b} was achieved by reducing background impurities.  A recent analysis by \cite{Ahn2022} points out that the new record brings the measured transport gap to within $40\%$ of an ideal theoretical gap that includes only finite thickness (see Sec.~\ref{sec_numerical_methods} for a discussion of finite thickness in numerics). 

Confinement, pressure, and in-plane electric fields have also been explored as sample design parameters. Survival of the 5/2 state under lateral confinement in quantum point contacts (QPCs) is important for edge tunneling and braiding experiments.  It was found that various confinement protocols can drive the formation of an even denominator FQH effect in the second LL~\cite{Fu2016}. But the direct application of a weak in-plane electric field in the Corbino geometry~\cite{Zhu2018} did not show enhancement of the strength of the 5/2 state as expected~\cite{Tylan-Tyler2017}.  Pressure and strain can play a role in orientation of the nearby nematic phase~\cite{Koduvayur2011} discussed below.  They were found to reduce the gap at 5/2 and drive a transition to the nematic phase~\cite{Samkharadze2016, Schreiber2020, Schreiber2018}. 

\subsubsection{In-plane magnetic field and sample tilt impacts the bulk gap} 
\label{sec_tilt}

An in-plane field increases Zeeman energy in comparison to the Coulomb energy thus allowing a tuning of the ratio of these two energies.  Increasing Zeeman energy would, in principal, favor states with higher spin polarization.  Sample tilt in a fixed magnetic field was used to explore this ratio in the context of the important question of the polarization of the 5/2 state (see Sec.~\ref{sec:spinpolarization}). 

Sample tilt simultaneously alters several energy scales at once.  For example, an in-plane field imposed by sample tilt can also alter the interaction by impacting finite thickness effects thus lowering the energy of competing ground states at 5/2~\cite{Peterson2008, Peterson2008a}.  As a result, experiments find differing behavior depending upon width, density, and tilt angle.  For example, increasing tilt lowers the transport gap in some samples~\cite{Dean2008, Dean-gap-5/2, Eisenstein1988, Eisenstein1990, Xia2010, Zhang2010}, while others find weak dependence \cite{Csathy2005, Wang2020} or even an enhancement~\cite{Liu2012}.  

Sample tilt was found to introduce a competing anisotropic phase~\cite{Du1999,  Lilly1999, Pan1999}. The anisotropic phase is believed to be a compressible nematic (stripe) phase~\cite{Fradkin2010, Koulakov1996, Moessner1996, CDW1999, Haldane-Pf2000, Schreiber2020}.  The general trend found to date~\cite{Csathy2005, Dean2008, Dean-gap-5/2, Eisenstein1988, Eisenstein1990, Liu2012, Wang2020, Xia2010, Zhang2010} shows that increased tilt angles induce a transition from an isotropic incompressible state, to an anisotropic compressible state.  For even larger tilt angles the anisotropy is observed to diminish with no transport gap observed.

Concomitant anisotropy and incompressibility have been reported as well.  In addition to evidence reported in standard transport tilt experiments~\cite{Liu2013, Xia2011}, there is also evidence from optical methods~\cite{Levy2016} and non-linear transport in the Corbino geometry~\cite{Bennaceur2018} that there is concomitant FQH effect and charge density wave order even in the absence of tilt.  Observation of FQH effect requires a gap.  But charge density waves (assuming they cause anisotropy) tend to lead to compressibility.  Concomitant FQH effect and anisotropy (and/or density wave order) is therefore at odds with the idea that the 5/2 FQH effect derives from a spatially uniform incompressible quantum Hall liquid. Observations of coexisting orders could also derive from domain formation~\cite{WY2016}. 

\subsubsection{Optical probes of the bulk gap} 

An incompressible FQH state requires a bulk collective-mode gap at all wavevectors.  Optical studies of FQH states can be used to probe the bulk gap at many different wavevectors, not just the large wavevectors assigned to the transport gap.  Since optical wavelengths are much smaller than the magnetic length, optical probes would be expected to create nearly zero-wavevector excitations.  But the breakdown of translational invariance through, e.g., disorder, allows access to other wavevectors~\cite{Sarma1997}.  

Several optics experiments at 5/2 reveal a uniform incompressible state as expected from transport experiments~\cite{Du2019, Levy2016, Rhone2011, Stern-optical, Wurstbauer2013}. [Note, however, that there is evidence of domain formation \cite{Rhone2011}.]  \cite{Wurstbauer2013} report evidence of gapped low-lying excitations at 5/2 using resonant inelastic light scattering.  The excitations observed here are reported to be consistent with spin-conserving roton-like neutral modes (see Sec.~\ref{sec:spinpolarization}).  These results are also consistent with recent resonant inelastic light scattering experiments where the intra-LL plasmon spectrum characteristic of a nematic phase was observed to have a pronounced minimum at 5/2~\cite{Du2019}. \cite{Du2019} argued that this was evidence for the strength of the uniform liquid paired state.  Optical probes have also been used to infer spin polarization as discussed in the following section. 

\subsection{Surface acoustic waves, geometric resonance, transport, light scattering, and Knight shift: spin polarization}
\label{sec:spinpolarization}

Antisymmetry of the total electron wave function requires that odd $l$-wave chiral BCS pairing of composite fermions be in a symmetric spin state, e.g., full spin polarization.  The Pfaffian wave function and other non-Abelian wave functions considered in the literature (see Table~\ref{tab:summary}) usually assume full spin polarization. Full spin polarization is therefore believed to be an important condition for non-Abelian quasiparticle candidate states that have received the most attention in the literature.  Although one can, in principle, construct non-Abelian states without full spin polarization, see, e.g., \cite{Barkeshli2010-2, Barkeshli2010-1, wide-well, YangRezayi2008}. The following section reviews experimental findings regarding the assumption of full spin polarization.  

\subsubsection{Surface acoustic waves and geometric resonance}

Pairing of composite fermions in a fully polarized state requires a fully polarized composite fermion Fermi sea~\cite{Read-Green} that can be thought of as originating from a composite fermion Cooper instability~\cite{Scarola_nature}.  Surface acoustic wave measurements were used to deduce the existence of a composite fermion Fermi sea in the half-filled lowest LL ($\nu=1/2$)~\cite{SAW-half} and a later measurement \cite{Willett2002} at 5/2 revealed similar Fermi surface properties, although at 5/2 the temperature had to be tuned to be high enough to leave behind a gapless composite fermion Fermi sea at 5/2.  Subsequent geometric resonance measurements~ \cite{polarized-CF, Mueed2017} went further to reveal evidence for a fully polarized Fermi sea near 5/2.  These experiments thus provide evidence for an underlying polarized composite fermion Fermi sea that sets the stage for an instability to a fully polarized incompressible paired state at low temperatures at 5/2. 

\subsubsection{Transport}

We now turn to transport experiments designed to infer the polarization of the gapped 5/2 state.  Tilted field transport gap measurements were the first to be used to infer the ground state polarization~\cite{Csathy2005, Dean2008, Dean-gap-5/2, Eisenstein1988, Eisenstein1990, Liu2012, Wang2020, Xia2010, Zhang2010}.  But as discussed above, the presence of competing ground states with tilt complicates these interpretations.  Early studies found a collapse of the transport gap upon tilt which was interpreted as evidence for an unpolarized (or partially polarized) state~\cite{Eisenstein1988, Eisenstein1990}.  However, the subsequent observation of a competing anisotropic phase at higher tilt~\cite{Lilly1999, Pan1999} shows that the reduction of the 5/2 transport gap was most likely due to competing phases for these parameters \cite{Dean2008,Friess2014}.  

Density tuning, as opposed to sample tilt, was used as an alternate route to explore ground state polarization as inferred from the transport gap.  Density and magnetic field dependence over a wide range initially showed~\cite{Pan2001} no transition (in support of a full polarization). But subsequent studies~\cite{Pan2014,Samkharadze} found evidence consistent with a spin transition from a fully polarized state at high density to an unpolarized state at low densities where LL mixing may play a strong role. 

\subsubsection{Light scattering}
\label{sec:raman}

Light scattering has been used to probe the polarization and excitations of the 5/2 state. Polarization resolved photoluminescence spectroscopy at 5/2 exhibited a sharp drop near 5/2 which was initially interpreted as evidence for an unpolarized state \cite{Stern-optical}.  And a combination of elastic and inelastic resonant scattering reveals a spin-excitation continuum that suggests competing phases and possibly polarized/unpolarized domains slightly away from filling 5/2 \cite{Rhone2011}.  But later resonant inelastic light scattering revealed spin waves only very close to filling 5/2 \cite{Du2019, Levy2016, Wurstbauer2013} that were argued to reveal full (or nearly full) spin polarization. 

\begin{figure}[htb]
\centering
\includegraphics[width=0.45\textwidth]{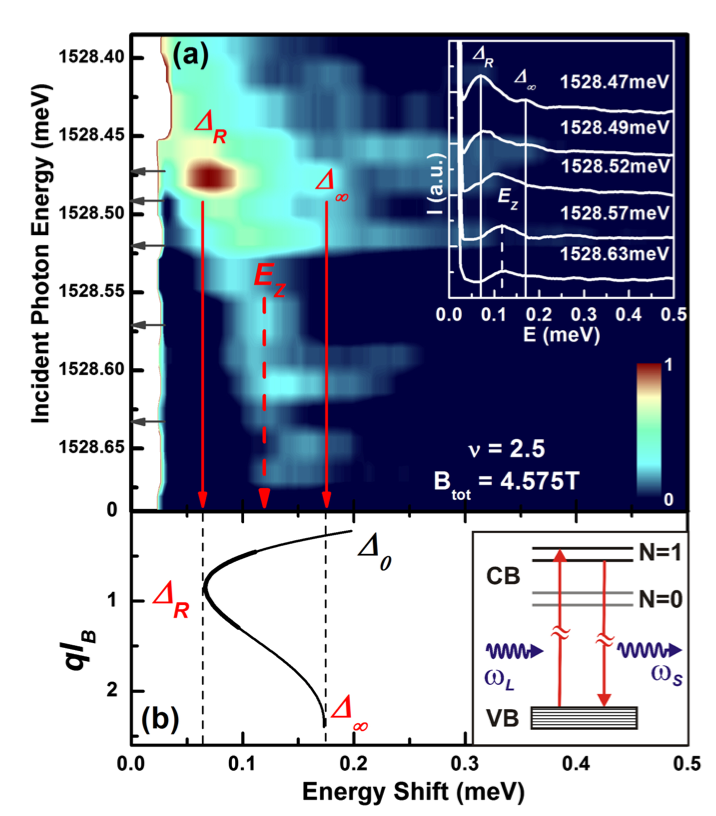}
\caption{(a) Resonant inelastic light scattering intensities measured at $\nu=5/2$ as a function of the incoming photon energy.  The inset shows spectra at energies marked by horizontal arrows in the color plot. $E_Z$ labels the Zeeman energy to show evidence for spin-wave modes.  (b)  Empirical wavevector ($q\ell_0$) dispersion based on the observed resonant inelastic light scattering mode energies showing at least one deep roton minimum at energy $\Delta_R$.  $\Delta_{0}$ and $\Delta_{\infty}$ label the small and large wavevector gaps, respectively.  The inset shows a schematic of the optical transitions between the valence band (VB), conduction band (CB), and the LLs labelled by $N$.  $\omega_L$ and  $\omega_S$ label the incoming and scattered photon frequencies, respectively. The figure is taken from \cite{Wurstbauer2013} with permission.}
\label{fig:lightscatteringexperiment}
\end{figure}

Figure~\ref{fig:lightscatteringexperiment} shows resonant inelastic light scattering data from \cite{Wurstbauer2013} performed at 5/2.  Sample tilt is used to impart finite momentum transfer in photon backscattering. [For a review of these methods see, e.g., \cite{Sarma1997}.] Figure~\ref{fig:lightscatteringexperiment} reveals both a magnetoexciton collective mode (see Sec.~\ref{sec:numerics}) and small wavevector spin wave modes excited near the Zeeman energy.  The coexistence of these modes was interpreted to be consistent with a fully polarized state with a magnetoexciton, as expected for polarized candidate states such as the Pfaffian and anti-Pfaffian. A gapless mode was also found at fillings slightly away from 5/2, consistent with other optical experiments \cite{Rhone2011}.  These and other light scattering data therefore provide evidence for a gapped fully spin polarized state at fillings very close to 5/2 but a gapless mode (possibly due to domain formation or competing phases) slight away from filling 5/2~\cite{Du2019, Levy2016, Rhone2011, Stern-optical, Wurstbauer2013}.

\subsubsection{Knight shift}

Strong evidence for full spin polarization at 5/2 also came from resistively detected nuclear magnetic resonance experiments~\cite{Stern2012, Tiemann2012}.  These experiments rely on the hyperfine interaction between the electrons in the GaAs quantum well and the atomic nuclei.  Shifts in the nuclear resonance frequency are used to infer the polarization of the electron gas (Knight shift).

\begin{figure*}[htb]
\centering
\includegraphics[width=6in]{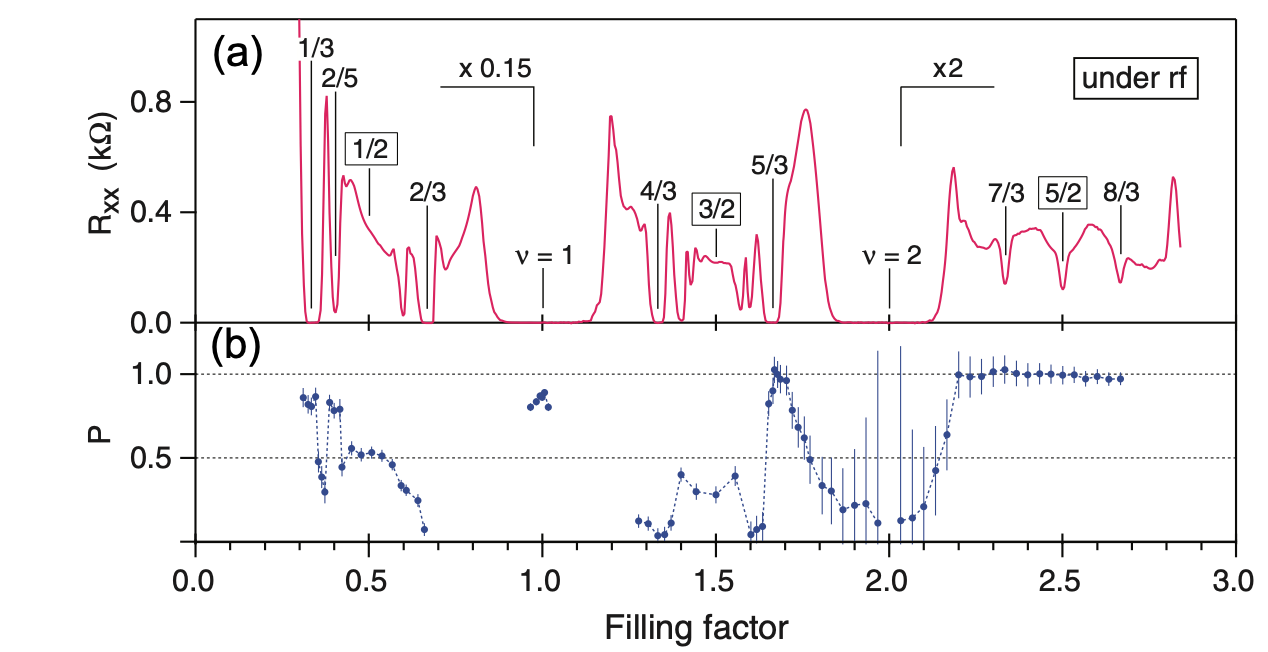}
\caption{(a) Longitudinal Hall resistance measured as a function of total filling factor in the presence of a radio frequency (rf) driving field.  (b)  The electron spin polarization ($P$) deduced from restively detected nuclear magnetic resonance plotted as a function of total filling factor.  The figure is taken from \cite{Tiemann2012}.}
\label{fig:nmrexperiment}
\end{figure*}

Figure~\ref{fig:nmrexperiment} shows data from \cite{Tiemann2012}. The data show that the ferromagnets expected at fillings $\nu=1$ and $\nu=1/3$ are nearly fully polarized whereas the data at and near $\nu=5/2$ reveal full polarization (within error bars). 
These and other Knight shift data were interpreted as evidence for full spin polarization~\cite{Stern2012, Tiemann2012} although some theory work suggests interpretation of full polarization is not always justified~\cite{Chesi2008}. 

\subsection{Tunneling with QPCs and SETs: quasiparticle charge and edge exponents}

Candidate ground states at 5/2 all have dominant quasiparticle charges of $e/4$ (see Table~\ref{tab:summary}).  Experiments conclusively identifying bulk quasiparticle charge of $e/4$ are therefore crucial in confirming that the ground state at 5/2 is a paired state, a necessary condition before detecting their non-Abelian nature.  Edge theories of various proposed states also rely on assumptions regarding quasiparticle charge.  Many different edge tunneling exponents are predicted from these theories (Table~\ref{tab:summary}).  The tunneling exponent, $g$, is inferred from tunneling conductance measurements (see Sec.~\ref{sec:propanyons}). The following reviews experimental determination of quasiparticle charge and tunneling exponents with QPCs and SETs at filling 5/2. 

\subsubsection{QPC tunneling}

Shot noise in tunneling across a QPC can be used to determine quasiparticle charge of a FQH liquid \cite{Yang-Girvin-book}.  Initial work \cite{Dolev2008} found evidence for $e/4$, although subsequent work \cite{Dolev2010} finds the charge to be higher than expected when in a regime of very weak backscattering probability and sufficiently small excitation energies. Here the measured value of charge was found to vary with temperature and applied voltage. This puzzling result might be explained by  the transport properties at low energies being dominated by  agglomerate excitations with charge $e/2$, rather than the $e/4$ quasiparticles~\cite{Carrega-noise}. These two contributions may be distinguished by  finite-frequency noise measurements~\cite{Carrega2012}.

The quasiparticle charge can also be inferred from the detailed temperature and/or bias votage dependence of QPC tunneling conductance.  Here two parameters, the quasiparticle charge and the tunneling exponent are extracted.  Several works~\cite{Baer2014,Fu2016,Radu2008, Venkatachalam_nature} extracted a quasiparticle charge consistent with or below $e/4$.  The non-universality of the observed charge at 5/2 is consistent with observations at other fractions \cite{Dima-fractional}.

\subsubsection{SETs}

SETs offered a method to extract local quasiparticle charge without relying on edge physics \cite{Venkatachalam_nature}.  Here the inverse compressibility in a local region is measured and the quasiparticle charge is inferred from gate-tuned jumps.  A value of $e/4$ was measured after disorder averaging.  These experiments offer the most direct measurement of $e/4$ because they are local and avoid edge tunneling.  

\subsubsection{Edge exponents from tunneling conductance}

Tunneling exponents extracted from tunneling conductance measurements can, unlike quasiparticle charge, be used to discern between various candidate states in Table~\ref{tab:summary}. The tunneling exponent (Sec.~\ref{sec:propanyons}) for the Majorana-gapped edge-reconstructed Pfaffian state is expected to be 0.5~\cite{Overbosch2008}. In weak backscattering tunneling conductance measurements~\cite{Baer2014, Fu2016, Lin2012, Radu2008} the tunneling exponent was found to be smaller than $0.4$, which is consistent with Abelian states.  But confinement was found to alter the value of the tunneling exponents \cite{Fu2016}.  This is consistent with more recent work which finds tunneling exponents that are incompatible even with established theories of the lowest LL, e.g., the 1/3 state \cite{Hennel2018} [see  also Section VIII in ~\cite{Dima-fractional} for further discussion].  

\subsection{Noise in QPCs: upstream neutral mode}

Some candidates for the $5/2$ state possess topologically protected upstream neutral modes. These modes do not carry charge, so it is difficult to detect them in an electrical transport experiments. However, they carry energy and can generate excess upstream noise.

Noise in quantum point contact experiments have been used to detect the presence of upstream neutral modes at filling 5/2~\cite{Bid2010, Dolev2011, Gross2012}. 
The detection of upstream neutral modes would appear to cast serious doubt on the Pfaffian state in real samples since only downstream modes are  expected (see Fig.~\ref{fig:edge}). On the other hand, it has been pointed out that edge reconstruction would allow an upstream neutral mode in a Pfaffian state~\cite{Bid2010, Overbosch2008}.  

\cite{Dolev2011} went on to explore the nature of the upstream neutral modes.  They compared those found at 5/2 to those at 7/3 and 8/3 on the same device. They concluded that the neutral upstream modes are not due to edge reconstruction.  Their conclusion casts doubt on the possibility of a Pfaffian ground state but supports an anti-Pfaffian state instead since it does posses an upstream neutral mode in the absence of edge reconstruction (Fig.~\ref{fig:edge}).

\subsection{Interferometry: quasiparticle charge and braiding}

Interferometry in the FQH regime relies on surface patterned gates to define multiple QPCs that guide edge currents in closed or open loops depending on gate voltages.  The loops encircle Aharonov-Bohm fluxes to probe the accumulated phase of the wave function.  The response of the device therefore depends on both the quasiparticle charge and the area enclosed.  As a result, interference experiments can be used to measure both quasiparticle charge and quasiparticle braiding statistics.  In this section we briefly review interferometry measurements of quasiparticle charge and braiding. More detailed discussions can be found in the reviews~\cite{Carrega2021,Dima-fractional, Willett-review}.

\subsubsection{Interferometry and charge}

The Aharonov-Bohm phase depends on the quasiparticle charge.  Interferometry measurements near 5/2 observed oscillations as a function of Aharonov-Bohm flux.  The flux was tuned in two ways: magnetic field strength or area (tuned by a side gate).  The observed oscillation patterns consistent with quasiparticle charge of both $e/4$ \emph{and} $e/2$ were found \cite{Willett2009,Willett2010-FPI}, but  they coexist only at low temperatures. As temperature was increased, oscillations consistent with just $e/2$ charge survive. This was anticipated by \cite{Yang-Pf2008}, who found the coherence length of the $e/4$ quasiparticles is much shorter than that of the $e/2$ quasiparticles, because the former involves the slow neutral mode velocity while the latter does not. As a result increasing temperature (which reduces coherence length) suppresses interference signal from the $e/4$ charge first.

\subsubsection{Interferometry and braiding}
\label{sec:FPI-braiding}

We now review interferometry measurements that provide evidence for non-Abelian braiding statistics of quasiparticles in the $5/2$ state.  Interferometry experiments are expected to reveal topological properties of anyons through anomalies in the Ahoronov-Bohm effect~\cite{Bonderson2006, DasSarma2005, Chamon1997, Fradkin1998, Das-Sarma, Stern2006}.  A topological even-odd effect in interferometry was expected to lead to smoking-gun evidence of non-Abelian order (if present) derived from fusion rules summarized in Fig.~\ref{tab:Ising-CFT}.  Nevertheless, it was later argued that some Abelian orders can also demonstrate the same effect under certain circumstances~\cite{Stern_PRB2010}.

Several works went on to report evidence for non-Abelian excitations in interference experiments~\cite{An2011, Willett2013, Willett2019}.  In these experiments resistance oscillations as a function of magnetic field were observed (similar to the oscillations used to infer quasiparticle charge).  Here, however, periods consistent with the magnetic field needed to add one or more non-Abelian quasiparticles were observed as phase slips.  The case for such oscillations were strengthened by controls done at filling 7/3 where only Abelian quasiparticles were expected.  \cite{Willett-review} summarizes next steps and hurdles for operation of a device as described in \cite{DasSarma2005}.  These include: phase stability, better control over device area, control of state changing current with time, and reproducibility in interference oscillations.

\subsection{Summary of early experimental findings}

We summarize early experimental support for assumptions underlying the possibility of candidate states hosting non-Abelian quasiparticles at filling 5/2.  A culmination of experimental work finds that the bulk energy gaps measured in both optics and transport are consistent with theoretical models that account for finite thickness and LL mixing in a spin polarized state. It is expected that more realistic modelling that includes disorder would bring theory and experiment into better agreement.  Furthermore, the culmination of evidence from both theory and experiment support the full polarization of the electrons in the half filled second LL, although there is some evidence of domain formation.   All experiments tend to support a measured quasiparticle charge of $e/4$ with some evidence for an additional quasiparticle with charge $e/2$.  Whereas results from quasiparticle edge tunneling, braiding, and shot noise have, so far, not been able to conclusively demonstrate sufficient conditions for non-Abelian quasiparticles.  

\section{Recent surprises and puzzles}
\label{sec:thermal-Hall}

The vast collection of experimental results reviewed in the previous section has provided invaluable insights into our understanding of the $5/2$ state. Meanwhile, the underlying topological order remains elusive. This challenge can, in principle, be met by measuring the thermal Hall conductance,
\begin{eqnarray} \label{eq:K-H}
K_H
= c\left(\frac{\pi^2 k_B^2 T}{3h}\right)
= c\kappa_0 T.
\end{eqnarray}
Here, $c=c_d-c_u$ is the chiral central charge of the CFT describing the edge of the sample~\cite{Cappelli_thermal, Kane_thermal, Read-Green}. Note that $c_d$ and $c_u$ are the total central charges for the downstream and upstream modes, respectively. When the edge of the quantum Hall system has counterpropagating edge modes, the interpretation of Eq.~\eqref{eq:K-H} becomes subtle. This issue will be discussed in Sec.~\ref{sec:partial}.

Two types of edge modes are relevant to the $5/2$ state. A Bose mode has central charge $1$ or $-1$. This is independent of whether the mode is charged or neutral. A Majorana fermion mode has central charge $1/2$ or $-1/2$. The positive/ negative value is adopted when the mode is downstream/ upstream in both cases. Therefore, each topological order in Table~\ref{tab:summary} has its predicted thermal Hall conductance at $\nu=5/2$ given by
\begin{eqnarray}
K_H=3+\frac{\mathcal{C}}{2}.
\end{eqnarray}
The first term on the right hand side originates from the three Bose modes that exist in any topological order describing the $5/2$ state. All of them are running downstream. Note that $K_H<0$ indicates that heat (or energy) flow along the edge in the opposite direction as the flow of charges, but the sign of $K_H$ cannot be resolved in the usual two-terminal setup.

\subsection{Experimental results}

\begin{figure} [htb]
\centering
\includegraphics[width=3.1in]{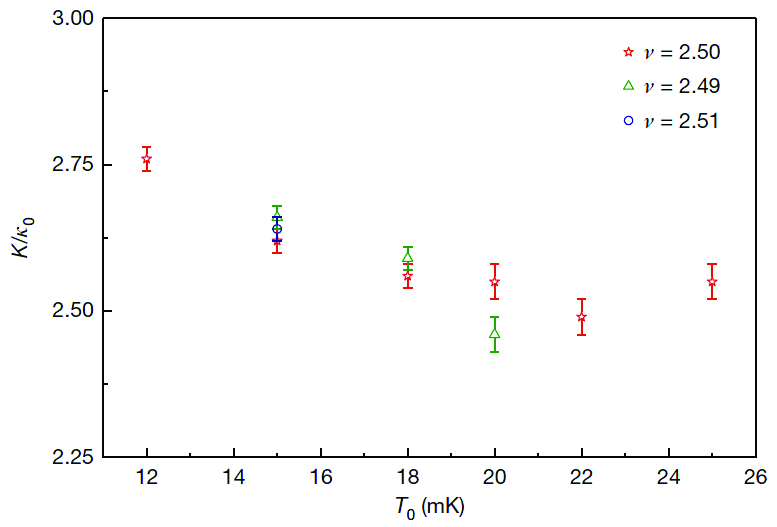}
\caption{Thermal conductance as a function of temperature for the $5/2$ state in a GaAs heterostructure at electron number density $\rho\approx 3.0\times 10^{10} \text{cm}^{-2}$. Data is taken from~\cite{Banerjee2018} with permission.}
\label{fig:th-Hall}
\end{figure}

The first set of experimental data of thermal conductance measurement in the $5/2$ state was reported recently in~\cite{Banerjee2018}. The data is shown in Fig.~\ref{fig:th-Hall}.

At temperature around $20$ mK, $K_H$ was nearly quantized at $2.5\kappa_0$. This result is very exciting as the half-integer quantized value for $K_H$ strongly supports the non-Abelian nature of the $5/2$ state. Meanwhile, the result remains puzzling. It does not agree with the theoretical predicted values, namely either $K_H=1.5\kappa_0$ for anti-Pfaffian state or $K_H=3.5\kappa_0$ for Pfaffian state. A more detailed analysis of the data also reveals a very rapid increase of $K_H$ at low temperature (below 18 mK). This growth is much faster than any other filling factors~\cite{Banerjee2017}. 

\subsection{Possible interpretations}

Both the unusual growth of $K_H$ at low temperature and its nearly quantized value at $2.5\kappa_0$ can be explained naturally if the sample has the PH-Pfaffian edge structure (see Sec.~\ref{sec:theory}). However, the PH-Pfaffian state has not received much numerical support as the Pfaffian or anti-Pfaffian state (see Sec.~\ref{sec:numerics}). The apparent tension between experimental data and existing numerical results motivated various proposals (or scenarios), which we now turn to.

\subsubsection{Bulk edge correspondence: PH-Pfaffian}

As already mentioned, the most straightforward interpretation is that the bulk possesses the PH-Pfaffian order. Then, one would need to understand how to reconcile with existing numerical studies, which largely ignored effects of disorder and LL mixing, although the latter has been treated perturbatively in some cases. It was first suggested that the PH-Pfaffian state may be stabilized by a combination of LL mixing and disorder effects~\cite{Zucker2016}. Meanwhile, the stabilization of PH-Pfaffian state by sufficiently strong LL mixing alone in uniform systems was also proposed~\cite{Milovanovic_LLM, Milovanovic2020}. This idea is corroborated by demonstrating explicitly the possible emergence of the PH-Pfaffian state in translationally and rotationally invariant systems~\cite{SMF2020}. In particular, more than one LL needs to be involved in stabilizing the PH-Pfaffian state~\cite{Milovanovic2021}. 

\subsubsection{Disorder induced Pf-APf domain walls}

While it is challenging to include disorder directly in microscopic numerical studies, its effect can be studied in a more phenomenological manner. In the limit of zero LL mixing such that particle hole symmetry is preserved, the Pfaffian and anti-Pfaffian states are degenerate at $\nu=5/2$. Disorder breaks the particle hole symmetry locally, and may induce Pfaffian and anti-Pfaffian domains. The domain walls are found to support gapless neutral (Majorana fermion) excitations, but have a charge gap~\cite{BMF2015,WY2016}. 

When the domain walls come close enough to each other, these neutral excitations can tunnel across the domain walls. The resulting percolation phase from the four copropagating Majorana fermions on each domain wall determines the thermal Hall conductance of the system. Using an effective network model, it was found that different phases can be realized in the system depending on the disorder strength~\cite{Lian2018, Mross2018, Wang2018}. In principle, there is a phase with $K_H=2.5\kappa_0$ and share the topological properties of the PH-Pfaffian state. Nevertheless, this phase is not favored for generic parameters in the network model analyzed by~\cite{Wang2018}. Its realization requires either the assumption that it is only a subdominant phase nucleated in the domain walls, or constraining the scattering events for Majorana fermion modes along domain walls by additional symmetry.

Using the technique of density matrix renormalization group, subsequent work studied the energetics of domain walls. It was originally concluded that the formation of domain wall is energetically unfavorable~\cite{Simon-domain}. However, the later work~\cite{Zhu-domain} suggested that the intrinsic electric dipole moment emerging at the interface not taken into account in~\cite{Simon-domain} can stabilize the Pfaffian and anti-Pfaffian domains under experimental conditions.

The domain states discussed here are quite similar to the stripe state of~\cite{WY2016} [proposed before the experiment~\cite{Banerjee2018}], where stripes (or domains) of Pfaffian and anti-Pfaffian states form spontaneously. Such a state restores particle-hole symmetry on average.

\subsubsection{Partial equilibration on anti-Pfaffian edges}
\label{sec:partial}

In the above two scenarios, one holds the view that the realistic sample has the PH-Pfaffian edge structure, although they have different bulk origins. This is reasonable because the value $2.5\kappa_0$ is only consistent with the PH-Pfaffian state (see Table~\ref{tab:summary}). At the same time, this conclusion actually relies on the assumption that the edge of the sample is fully equilibrated. When the edge has downstream modes only, this assumption is always satisfied independent of the system size. Then, the thermal Hall conductance should agree with the theoretical predicted value in Eq.~\eqref{eq:K-H}. 

The situation becomes more subtle if the edge has counterpropagating modes. In this case the downstream and upstream modes along the same edge originate from two different sources at different temperatures~\cite{Banerjee2018}. The distance required for a pair of counterpropagating edge modes to reach thermal equilibration is known as the thermal equilibration length, denoted as $\ell_{\rm eq}$. In order to achieve a complete thermal equilibration on the edge, the length of the edge, denoted as $L$, needs to satisfy $L\gg\ell_{\rm eq}$~\cite{Aharon2018, Banerjee2017}. 

The above subtlety gives rise to a possible reconciliation between the experimental data and the realization of anti-Pfaffian state in the sample (including the edge). When the edge is fully equilibrated, the anti-Pfaffian state has $K_H=1.5\kappa_0$. This is because the net central charge of the edge theory is given by $c=3-3(0.5)=1.5$. On the other hand, the two central charges of a pair of counterpropagating edge modes should add instead of subtracting each other if these two modes do not equilibrate. Hence, $K_H/\kappa_0$ can be as large as $3+3(0.5)=4.5$ in the anti-Pfaffian state, if all upstream modes do not thermally equilibrate with the downstream modes. For the measured value $K_H=2.5\kappa_0$, it can be explained by partial thermal equilibration on the anti-Pfaffian edge. For the details of the mechanism, various proposals were introduced and debated~\cite{Mulligan2020, Dima_comment2018, Ken2019, Simon_PRB2018, Simon-reply, Simon2020}. 

\subsection{Recent and ongoing developments}

In an attempt to distinguish the above scenarios, several subsequent experimental and theoretical studies have been carried out.

\subsubsection{Temperature dependence of $\ell_{\rm eq}$}

Besides the value $2.5\kappa_0$, it is also important to explain the unusually rapid growth of $K_H$ at low temperature. An increase of $K_H$ at low temperature is actually reasonable because $\ell_{\rm eq}$ is temperature dependent. Most of the topological orders (with both downstream and upstream edge modes) have $\ell_{\rm eq}\sim T^{-2}$. For a sample edge with length $L$, the ratio $L/\ell_{\rm eq}$ decreases as $T$ is lowered. This can lead to a lack of thermal equilibration on the edge, and leads to an increase in $K_H$ at low temperature~\cite{Aharon2018, Banerjee2017}. 

The special structure of the PH-Pfaffian edge poses stringent constraints on possible interaction terms that couple the downstream Bose mode $\phi_c$ and the single upstream Majorana fermion mode $\psi$ (see Fig.~\ref{fig:edge}). The most relevant interaction is described by the operator,
\begin{eqnarray}
\hat{O}(x)=\eta(x)\partial_x\phi_c(\psi\partial_x\psi),
\end{eqnarray}
with $\eta(x)$ being a random variable in space. The scaling dimension of $\hat{O}(x)$ is $\Delta=3$, which is larger than the usual operators that couple two counterpropagating Bose modes, or a downstream Bose mode and a pair of different upstream Majorana fermion modes. In the language of renormalization group, $\hat{O}(x)$ describes a more irrelevant process and leads to $\ell_{\rm eq}\sim T^{-4}$ in the PH-Pfaffian edge~\cite{Ken-thermal}. This exceptional dependence in temperature for the thermal equilibration length may explain the unusual rapid growth of $K_H$ at low temperature, if the system has the PH-Pfaffian edge structure.

\subsubsection{Interface experiments}

\begin{figure*} [htb]
\centering
\includegraphics[width=6.0in]{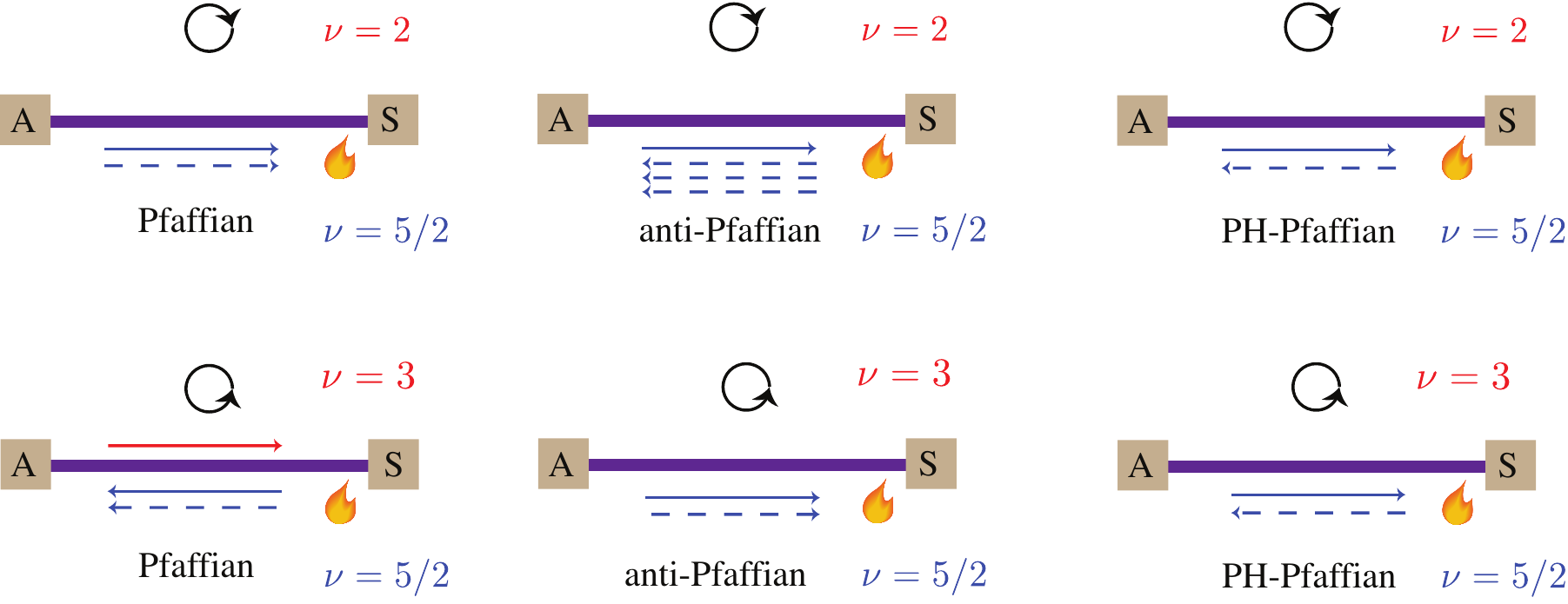}
\caption{Schematic diagram for the freely propagating (gapless) modes along the $5/2-2$ and $5/2-3$ interfaces when the $5/2$ state has Pfaffian, anti-Pfaffian, or PH-Pfaffian edge. The solid lines and dashed lines label Bose charge modes and neutral Majorana fermion modes, respectively. The square with label A and S represent respectively the amplifier and source. The effective interface length is defined as the distance between A and S. A hot spot (denoted as the fire mark) is formed at the back of the source when a dc current is injected at the interface from $S$. Depending on the gapless modes' chirality, excess noise may be measured at the amplifier or not. Note that the direction of magnetic field was reversed in the experiment when switching from the $5/2-2$ to the $5/2-3$ interface, so that the positions of the source and amplifier needed not be interchanged.}
\label{fig:interface}
\end{figure*}

One of the main differences among possible candidates for the $5/2$ state is the number and chirality of neutral modes on the edge. Thus, it is desirable to probe them more directly. By placing a region with $\nu=5/2$ close to another region with $\nu=2$ or $\nu=3$, a resulting $5/2-2$ or $5/2-3$ interface is formed. Some of the original edge modes are gapped out or become localized along the interface. For Pfaffian, anti-Pfaffian, and PH-Pfaffian edges, the modes that remain freely propagating along the interfaces are illustrated in Fig.~\ref{fig:interface}. An important consequence is that these gapless modes dominate both charge and energy transport along the interface.

By injecting a dc source current at the interface, a hot spot will be created at the back of the source. Relatively short interfaces (28 $\mu$m, 38 $\mu$m, 48 $\mu$m, and 58 $\mu$m) were employed in the experiment that realized thermally unequilibrated transport. In this regime, shot noise will be detected at the amplifier when there are gapless modes propagating away from the source to the amplifier. In the experiment, shot noise was observed in both $5/2-2$ and $5/2-3$ interfaces~\cite{Heiblum-interface}. If one neglects the possibility of reconstruction along the interface, the experimental result is only achievable if the original $5/2$ state has the PH-Pfaffian edge. This is because the PH-Pfaffian edge is the only one that remains unchanged under particle hole transformation. Furthermore, it has been theoretically proposed that electrical conductance measurement  in a device that simultaneously contains both $5/2-2$ and $5/2-3$ interfaces could also identify the Pfaffian, anti-Pfaffian, and PH-Pfaffian edges uniquely~\cite{Yutushui2022}.

Another subsequent experiment was performed to measure the thermal conductances of the $5/2-2$ and $5/2-3$ interfaces~\cite{Heiblum-interface2}. In order to achieve thermal equilibration for a proper thermal conductance measurement, a much longer interface with length 160\textmu m was used. The measured values for the $5/2-2$ and $5/2-3$ interfaces were $(0.55\pm 0.02)\kappa_0$ and $(0.53\pm 0.02)\kappa_0$, respectively. For Pfaffian and anti-Pfaffian edges, either the resulting $5/2-2$ or $5/2-3$ interface will have a pair of copropagating Bose and Majorana fermion modes. These two modes are always equilibrated, so $K_H/\kappa_0=1.5$ is expected. Hence, the observed values in experiment favors the PH-Pfaffian edge in the original $5/2$ state.

\subsection{Summary of recent developements}

To summarize, the results from recent experiments have provided increasing support for the realization of PH-Pfaffian edge in the $5/2$ state. This edge structure is also consistent with previous results from tunneling experiments and the observation of upstream neutral modes. On one hand, the bulk-edge correspondence suggests that the bulk of the sample should be also described by the PH-Pfaffian state. On the other hand, the domain-wall picture hints at a possible difference between the bulk and the edge of the sample. Therefore, the underlying nature of the $5/2$ state is still an open problem.

\section{Future directions} \label{sec:future}

In this section, some possible future research directions for the $5/2$ state are listed.

\subsection{Noise measurements}

Besides the thermal conductance and interface experiments reviewed in the previous section, there are some recent theoretical proposals to identify the Pfaffian, anti-Pfaffian, and PH-Pfaffian edges via noise measurements. The corresponding results can further strengthen the support for or challenge the realization of PH-Pfaffian edges in real samples.

\subsubsection{Shot noise in dc transport}

It has been proposed that the combination of thermal conductance and shot noise measurements in dc transport can uniquely distinguish between the Pfaffian, anti-Pfaffian, and PH-Pfaffian edges. In addition to the edge modes coming from the second LL as shown in Fig.~\ref{fig:edge}, all states at $\nu=5/2$ have two additional edge modes originating from the lowest LL. 

Depending on the sample edge length, it is possible that edge modes from the second LL equilibrate among themselves, but not with those from the lowest LL~\cite{Ken2019}. In this scenario, both anti-Pfaffian and PH-Pfaffian edges will show $K_H=2.5\kappa_0$. However, the predicted shot noises in dc measurement are distinct for these two edges. Only the anti-Pfaffian edge will generate a nonvanishing shot noise~\cite{Park2020}. For more details of the topological classification of noises in different types of edges, see~\cite{Spanslatt2019}. For the Pfaffian edge, it has no upstream modes. Thus, $K_H=3.5\kappa_0$ independent of the sample edge length. At this moment, there is no proposal for the reconciliation between the thermal conductance data with the Pfaffian edge.

\subsubsection{Thermal tunneling noise across a QPC}

The scaling dimension of quasiparticles probed in tunneling experiments may help identify the edge structure of the $5/2$ state. However, existing results depend sensitively on the device configuration and lead to ambiguous interpretations.  It has been suggested that such an ambiguity could be removed by measuring the thermal tunneling noise across the QPC~\cite{Schiller2022}. This noise is generated due to quasiparticle tunneling, and should be defined as the excess noise away from the usual Nyquist-Johnson noise. In particular, the associated Fano factor will show different dependencies on the temperatures of the two edges and the voltage bias across the QPC, when the scaling dimension of the tunneling quasiparticle is changed. This feature allows one to distinguish between the anti-Pfaffian and PH-Pfaffian edges.

\subsection{Local power measurement in multi-terminal thermal conductance experiment}

The thermal conductance data reviewed in Sec.~\ref{sec:thermal-Hall} was taken from a two-terminal setup, which cannot resolve the sign of $K_H$. Neither can it give a universal value of $|K_H|$ when the edge is not fully equilibrated. This leads to ambiguous interpretation of the experimental result. It is important to develop techniques for measuring $K_H$ reliably, that should work independent of the equilibration on the sample edge.

In a very recent work, a multi-terminal measurement on thermal conductance was performed~\cite{Heiblum-thermal}. Using the new setup, the power carried by the downstream and upstream edge modes can be measured separately. From this, $K_H$ can be deduced properly from $K_H=K_d-K_u$, where $K_d$ and $K_u$ denote the thermal conductance for the downstream and upstream modes, respectively. At $\nu=2/3$, the new setup gives $K_H=(0.04\pm 0.03)\kappa_0$, which agrees with the theoretical predicted value, $K_H=0$ for an equilibrated edge at $\nu=2/3$. It would be natural to generalize the same kind of measurement to the $5/2$ state. The result can provide another piece of information to help  differentiate between the PH-Pfaffian and partially equilibrated anti-Pfaffian edges.

\subsection{Other interferometry experiments}

Direct measurement of braiding statistics of quasiparticles is an ideal  probe of the topological order of a FQH liquid. While the Fabry-P\'{e}rot interferometer has become a commonly used device for detecting non-Abelian statistics, there are some caveats in this technique which were discussed in Sec.~\ref{sec:FPI-braiding}.  The topological even-odd effect was originally proposed as a smoking-gun signal of non-Abelian nature of the 5/2 state~\cite{Bonderson2006, Stern2006}. However, it was later discovered that the Abelian 331 state~\cite{Stern_PRB2010}, and all other Abelian states (except the $K=8$ state) in the 16-fold way can mimic the same effect if the two types of $e/4$ anyons exhibit flavor symmetry~\cite{16-fold}.   It will be important to include these caveats in future experimental designs to better delineate expected order at 5/2.

A recent breakthrough on interferometry at $\nu=1/3$ in the lowest LL has significant implications for detection of non-Abelian excitations at $\nu=5/2$ \cite{Nakamura-FPI}.  In \cite{Nakamura2020}, Fabry-P\'{e}rot interferometry was used to directly observe Abelian anyon statistics at $\nu=1/3$.  Normally, such experiments are done in a regime where the strong Coulomb interaction between edge states and the bulk of the interferometer complicates interpretation of signals of anyonic phases.  But samples used in \cite{Nakamura2020} incorporate epitaxially grown layers used to screen (and therefore soften) these interactions.  The net result was a robust Aharonov–Bohm interference in the fractional quantum regime showing clear phase slips consistent with Abelian anyon fractional statistics.  Application of these seminal developments to 
$\nu=5/2$ is a logical next step. 

\subsubsection{Mach-Zehnder interferometry}

Issues in discerning topological orders can be circumvented by employing the Mach-Zehnder interferometer ~\cite{MZ-2003, KT2006, Ponomarenko2007, Ponomarenko2010}. Previous work discovered that the corresponding signatures (i.e., tunneling current and Fano factor) for each individual state in the 16-fold way are different~\cite{Feldman2006, KT_noise, 16-fold, Chenjie2010, Guang2015, Zucker2016}. Although it is much harder to fabricate an Mach-Zehnder interferometer than a Fabry-P\'{e}rot interferometer, the former was employed to yield evidence of fractional statistics of anyons in the FQH state at $\nu=2/5$~\cite{Heiblum-MZI}. It will be highly desirable to perform the same experiment for the 5/2 state.

\subsubsection{Non-Abelian anyon collider}

Besides the result from the standard Fabry-P\'{e}rot interferometry~\cite{Nakamura2020}, fractional statistics of anyons in the Laughlin state at $\nu=1/3$ was also observed in an anyon collider setup~\cite{Bartolomei, collider2016}. In a recent preprint~\cite{nA-collider}, it has been proposed that the latter technique may also differentiate among non-Abelian statistics of anyons in the Pfaffian, anti-Pfaffian, and PH-Pfaffian states. This may be achieved in future experiments. 

\subsection{Bulk probes: Polarized Raman scattering and thermal power}

While the bulk edge correspondence is a very powerful tool in studying topological phases, there is the possibility of its breakdown in realistic samples due to various complications at the edge as discussed earlier. In fact the edge physics of the Laughlin 1/3 state is still not yet completely understood \cite{RevModPhys.75.1449}, most likely due (at least in part) to edge reconstruction \cite{PhysRevLett.88.056802,PhysRevB.68.125307,PhysRevLett.91.036802}. It would thus be highly desirable to go beyond edge physics and probe the $5/2$ state (and other FQH states, Abelian or non-Abelian) in the bulk more directly in future experiments. We review some existing theoretical proposals below.

It was found that the graviton or magnetoroton in the Pfaffian state has a polarization $-2$~\cite{Liou-graviton}, just like the Laughlin state. Since the chirality of the graviton is reversed under particle hole transformation~\cite{Liou-graviton,Son-arxiv}, the associated graviton in the anti-Pfaffian state has a polarization $+2$, as seen in numerics~\cite{Haldane-Raman}. The situation becomes more subtle in the PH-Pfaffian state. Due to particle hole symmetry, it is expected that gravitons (if they are present) should come with both polarizations ($\pm 2$). 

In order to excite the graviton mode, a net angular momentum $+2$ or $-2$ needs to be transferred to the sample. This may be implemented by using Raman scattering with circularly polarized light~\cite{Golkar-JHEP}. Importantly, the absorption of a photon can occur only if the incoming photon has the correct polarization. For example, the incoming photon must have the polarization $-1$ in order to excite the graviton mode in the Pfaffian state~\cite{Liou-graviton}. On the other hand, the graviton mode in the anti-Pfaffian state requires an incoming photon with polarization $+1$. The emitted photon must have the opposite polarization such that the net angular momentum transfer is $\pm 2$ for anti-Pfaffian and Pfaffian respectively. For the PH-Pfaffian state, roughly equal absorption of both kinds of polarized photons is expected~\cite{Haldane-Raman, Son-Raman}. Suppose the sample contains Pfaffian and anti-Pfaffian domains, the same technique can also reveal this structure if the spatial resolution of the experiment is finer than the typical sizes of the domains. Since Raman scattering has already been performed in studying GMP modes in other quantum Hall states, it offers a huge potential in identifying the precise nature of the $5/2$ state in future experiment, which goes beyond the knowledge from edge probes.

The polarized Raman scattering discussed above directly probes the chirality of the FQH state (namely, whether it is electron-like or hole like), but not the nature of its quasiparticles. The latter, on the other hand, can be probed using bulk thermoelectric and thermodynamic measurements~\cite{PhysRevB.85.195107, PhysRevLett.102.176807, PhysRevLett.105.086801, PhysRevB.79.115317}. More specifically, such experiments can (potentially) measure the logarithm of the quantum dimension of the non-Abelian quasiparticles, which is the topological entropy they carry (Abelian quasiparticles have quantum dimension one and thus carry no topological entropy). Subsequent thermal power experiments have yielded encouraging but not yet definitive results \cite{PhysRevB.81.245319, PhysRevB.87.075302}. It would be highly desirable to continue this line of work.

\section{Conclusions}

The physics of the 5/2 FQH state, most likely the first and only non-Abelian state of matter identified in nature at the time of writing, is extremely rich. This richness, revealed through enormous experimental, theoretical, and numerical effort, underlies the decades-long debate over the precise nature of the ground and excited states.  Considerable progress has been made on all of these fronts, as we have reviewed here. The remaining unresolved puzzles that we have outlined in this chapter offer exciting opportunities for future research.

\section{Acknowledgments}

We would like to thank Ajit Balram, Matteo Carrega, Luca Chirolli, Gabor Cs\'{a}thy, Sankar Das Sarma, Dima Feldman, Moty Heiblum, Jainendra Jain, Thierry Jolicoeur, Sudhansu Mandal, Michael Manfra, Kwon Park, Edward Rezayi, Ganesh Sreejith, and Steve Simon for helpful suggestions and comments.  KKWM is supported by the Dirac postdoctoral fellowship in the National High Magnetic Field Laboratory. MRP acknowledges support from the Office of Research and Sponsored Programs at California State University Long Beach. VWS acknowledges support from AFOSR FA9550-18-1-0505,  FA2386-21-1-4081. KY acknowledges support from NSF grant No. DMR-1932796. The work of KKWM and KY was performed at the National High Magnetic Field Laboratory, which is supported by National Science Foundation Cooperative Agreement No. DMR-1644779, and the State of Florida.

\bibliographystyle{elsarticle-harv}
\bibliography{QHE-ref}

\end{document}